\documentclass[11pt,document,nofootinbib,superscriptaddress,onecolumn,preprintnumbers,balancelastpage]{article}
\pdfoutput=1
\usepackage{jheppub}
\usepackage{graphicx}
\usepackage{epstopdf}
\usepackage{dcolumn}
\usepackage{bm}
\usepackage{hyperref}
\usepackage{booktabs}
\usepackage[dvipsnames]{xcolor}
\usepackage{amsmath}
\usepackage{cancel}
\usepackage{xpatch}
\usepackage{mathtools}
\definecolor{c1}{rgb}{0.368417, 0.506779, 0.709798}
\definecolor{c2}{rgb}{0.880722, 0.611041, 0.142051}
\definecolor{c3}{rgb}{0.560181, 0.691569, 0.194885}
\definecolor{c4}{rgb}{0.922526, 0.385626, 0.209179}
\definecolor{c5}{rgb}{0.528488, 0.470624, 0.701351}
\definecolor{c6}{rgb}{0.772079, 0.431554, 0.102387}
\definecolor{c7}{rgb}{0.363898, 0.618501, 0.782349}
\definecolor{turq}{rgb}{0.181,0.638,0.594}
\definecolor{pink}{rgb}{1.000,0.54,0.8}
\definecolor{purple}{RGB}{155,100,155}
\definecolor{gray}{RGB}{128,128,128}
\definecolor{lightBlue}{RGB}{148,179,229}
\definecolor{lightRed}{RGB}{213,157,131}
\definecolor{violet}{RGB}{130,121,173}
\definecolor{gold}{RGB}{255,191,0}

\def\beq{\begin{align}}
\def\eeq{\end{align}}

\newcommand{\lsim}{ \mathop{}_{\textstyle \sim}^{\textstyle <} }

\makeatletter
\g@addto@macro\bfseries{\boldmath}
\makeatother

\title{
A Heavy QCD Axion and the Mirror World
}

\author[1,2,3]{David I. Dunsky}
\author[2,3]{Lawrence J. Hall}
\author[4,5,6,7,8]{Keisuke Harigaya}
\affiliation[1]{Center for Cosmology and Particle Physics, Department of Physics,
New York University, New York, NY 10003, USA}
\affiliation[2]{Department of Physics, University of California, Berkeley, California 94720, USA}
\affiliation[3]{Theoretical Physics Group, Lawrence Berkeley National Laboratory, Berkeley, California 94720, USA}
\affiliation[4]{Department of Physics, University of Chicago, Chicago, IL 60637, USA}
\affiliation[5]{Enrico Fermi Institute, University of Chicago, Chicago, IL 60637, USA}
\affiliation[6]{Kavli Institute for Cosmological Physics, University of Chicago, Chicago, IL 60637, USA}
\affiliation[7]{Kavli Institute for the Physics and Mathematics of the Universe (WPI),
The University of Tokyo Institutes for Advanced Study,
The University of Tokyo, Kashiwa, Chiba 277-8583, Japan}
\affiliation[8]{Theoretical Physics Department, CERN, Geneva, Switzerland}

\abstract{
We study the mirror world with dark matter arising from the thermal freeze-out of the lightest, stable mirror particle -- the mirror electron. The dark matter abundance is achieved for mirror electrons of mass 225 GeV, fixing the mirror electroweak scale near $10^8$~GeV.
This highly predictive scenario
is realized by an axion that acts as a portal between the two sectors through its coupling to the QCD and mirror QCD sectors.
The axion is more massive than the standard QCD axion due to additional contributions from mirror strong dynamics. Still, the strong CP problem is solved by this `heavy' axion due to the alignment of the QCD and mirror QCD potentials. 
Mirror entropy is transferred into the Standard Model sector via the axion portal, which alleviates overproduction of dark radiation from mirror glueball decays. This mirror scenario has a variety of signals: 
(1) primordial gravitational waves from the first-order mirror QCD phase transition occurring at a temperature near 35 GeV, (2) effects on large-scale structure  from dark matter self-interactions from mirror QED, (3) dark radiation affecting the cosmic microwave background, and (4) the rare kaon decay, $K^+ \rightarrow (\pi^+ + \rm{axion})$. The first two signals do not depend on any fundamental free parameters of the theory while the latter two depend on a single free parameter, the axion decay constant.
}

\date{\today}

\begin{document}
\maketitle
\flushbottom

\newpage

\section{Introduction}

A unified theory of nature may fragment into a low-energy effective theory having multiple sectors, each with its own gauge symmetry.  If the fermions fall into sets carrying gauge charges of only a single sector, then the lightest fermion(s) in each sector are stable. In the Standard Model (SM) sector these are the lightest neutrino, the electron, and the proton. The cosmological dark matter may be the lightest fermion of some other sector, the dark sector.  While this framework is plausible, there are many such schemes each with a variety of free parameters, so that it is hard to construct realistic theories that can be tested.  

A unique and highly predictive scheme is the mirror world, where a $Z_2$ symmetry leads to an exact copy of the sector describing the directly observed particles and interactions. Indeed, before it was known that our sector was described by gauge symmetry, this mirror world was introduced as a way to preserve parity in nature \cite{Lee:1956qn, Kobzarev:1966qya} as the $Z_2$ symmetry can also send $\bar{r} \rightarrow - \bar{r}$ and flip fermion chirality. Over successive decades, interest in the mirror world, and the possibility of mirror dark matter, has increased \cite{Okun:2006eb, Foot:2014mia}. Furthermore, with the discovery of the possibility of  kinetic mixing between photons of different $U(1)$ gauge groups \cite{Holdom:1985ag}, it was proposed that kinetic mixing could probe dark sectors \cite{Goldberg:1986nk}.  However, mirror dark matter, with the mirror spectrum identical to the SM spectrum, faces several challenges. Measurements of the cosmic microwave background require the mirror sector to be at a lower temperature than the SM sector, both from mirror neutrino contributions to dark radiation and from mirror baryon acoustic oscillations~\cite{Cyr-Racine:2012tfp,Bansal:2022qbi}. In addition, the $Z_2$ symmetry must be broken by interactions beyond the SM, or by initial conditions, so that mirror baryons have a larger cosmological abundance than SM baryons.  Furthermore, the mirror matter halos should not dissipate and collapse into disks like SM matter, and must avoid the constraints on self-interactions from observations of the Bullet cluster.

There is an alternative route to the mirror world. Beyond kinetic mixing,
there is one other interaction of dimension 4 that couples the two sectors, $|H|^2 |H'|^2$, where $H$ and $H'$ are the SM and mirror Higgs doublets.  This term is not excluded by any symmetry, and it changes everything! If it is absent, as discussed above, the two Higgs vevs are equal, $v'=v$, so that the mirror spectrum is identical to the SM spectrum.  Two cases emerge if this additional Higgs portal operator is significant.  In the first, the vacuum is determined at tree-level and has two phases \cite{Foot:2000tp}. The symmetric phase has the two vevs equal, and is excluded by Higgs physics at the LHC. In the asymmetric phase one vev vanishes, so that mirror matter is extremely light and it is unclear how it yields dark matter. A realistic possibility for mirror dark matter emerges from adding a soft $Z_2$ breaking Higgs interaction, giving a hierarchy of vevs with $v' > v$ \cite{Barbieri:2005ri}. 
The second case, more minimal and constraining, has the vacuum determined by radiative corrections, via the Higgs Parity mechanism~\cite{Hall:2018let}. This gives one vev much larger than the other, $v' \gg v$, with the ratio determined by the measured values of the Higgs boson mass, the top quark mass, and the QCD coupling constant. With current data the central value of $v'$ is $10^{12}$ GeV, with a $3 \sigma$ lower bound of $10^9$ GeV.

In a previous paper \cite{Dunsky:2019upk}, we explored this mirror world from Higgs Parity and found successful cosmologies with thermally produced mirror electron, $e'$,  dark matter. With a high reheat temperature after inflation, the freeze-out abundance of $e'$ leads to the observed dark matter if $v' = 10^8$ GeV, well below the predicted range of $v'$. Furthermore, after the mirror QCD phase transition the lightest mirror glueball decays to mirror photons leading to too much dark radiation. Remarkably, a realistic cosmology results because, with $v'$ of order $(10^9 - 10^{10})$ GeV, the $e'$ and dark radiation can be sufficiently diluted by entropy created by the decays of mirror neutrinos. This cosmology requires the top quark mass to be about $2 \sigma$ high. Alternatively, if only the SM sector is produced after inflation with a low reheat temperature, $e'$ dark matter can arise from freeze-in, via either the Higgs or kinetic mixing portal, for a wide range of $v'$.

In this paper, we study the above mirror world with the addition of a Peccei-Quinn (PQ) symmetry \cite{Peccei:1977hh,Peccei:1977ur} giving a KSVZ axion that is even under the $Z_2$ symmetry (for the case of a Weinberg-Wilczek axion see \cite{Berezhiani:2000gh}). This leads to a successful cosmology with $e'$ dark matter from freeze-out without the need for dilution. The new colored states carrying the PQ symmetry, the PQ quarks, have couplings with the Higgs boson that modify the Higgs Parity mechanism to allow $v' = 10^8$ GeV. Furthermore, the mirror glueball decays to axions that deposit their energy in the SM sector, avoiding over-production of dark radiation. 

While parity may solve the strong CP problem, it does not do so in the mirror world; it simply relates the non-zero strong CP parameters of the two sectors. A $Z_2$-even axion solves the strong CP problem and is much heavier than the conventional axion~\cite{Rubakov:1997vp,Fukuda:2015ana,Hook:2019qoh}, since the dominant contribution to its mass arises from mirror QCD which gets strong near 30 GeV. If such a heavy axion decays after neutrino decoupling it heats up the photon bath relative to the neutrinos, suppressing the neutrino contribution to the radiation energy. To avoid this the axion scale must be low, $f_a < 10^5$ GeV, and the axion mass high, of order 10 MeV.  Such a low-$f_a$, high-mass axion greatly ameliorates the quality problem of the PQ symmetry, which can be more easily understood as an accidental consequence of other symmetries.  

The mirror world has also been motivated as a setting for the Twin Higgs mechanism~\cite{Chacko:2005pe}, which improves the naturalness of the weak scale. In this case $v'$ is around the TeV scale, far below that predicted by the exact mirror world, so that soft breaking of $Z_2$ in the Higgs potential is required~\cite{Barbieri:2005ri}. Such soft breaking could arise from spontaneous breaking in some other sector. We also consider this possibility for obtaining $v' = 10^8$ GeV, needed for $e'$ dark matter from freeze-out, without the need of any Yukawa coupling between the PQ quarks and the Higgs boson. Refs.~\cite{Barbieri:2016zxn,Fukuda:2017ywn,Barbieri:2017opf} consider various dark matter candidates for the theories with soft and hard $Z_2$ breaking.

In addition to the SM parameters, the mirror world studied in this paper has three parameters relevant for cosmological and particle physics signals, $v', \epsilon$ and $f_a$. Without dilution from mirror neutrino decay, the $e'$ dark matter abundance fixes $v' = 10^8$ GeV. Thus, the effects of self-interactions on the dark matter halo ellipticity from mirror electromagnetism involve no free parameters.  Similarly, gravity waves from the mirror QCD phase transition involve no free parameters of the underlying particle theory, since the phase transition temperature can be computed. Furthermore, the gravity wave signal is not diluted by entropy generated from neutrino decay, as in \cite{Dunsky:2019upk}.  On the other hand, the predicted rate for $K \rightarrow \pi a$ depends on $f_a$. The amount of dark radiation is sensitive to axion physics and also depends on $f_a$; in addition it is sensitive to the electromagnetic anomaly of the PQ symmetry. Direct detection of $e'$ dark matter can occur via kinetic mixing and depends on $\epsilon$. If the SM is embedded in a unified theory at scale $v_G$, then $\epsilon$ depends sensitively on $v_G$, so that the direct detection signal is correlated with the proton decay rate~\cite{Dunsky:2019upk}.

We present the theory in section \ref{sec:theory}, including portal operators and the PQ sector, and discuss the spontaneous breaking of the $Z_2$ via Higgs vevs. 
In section \ref{sec:spectra} we elaborate on key aspects of particle phenomenology, including the spectrum of mirror fermions, properties of the axion and the lightest mirror glueball, and the generation of neutrino and mirror neutrino masses. The cosmological history is presented in section \ref{sec:cosmology}: thermal decoupling between the two sectors is studied, as well as the parameter space that yields the observed dark matter via mirror electron freeze-out. In section \ref{sec:Signals}, we present signals of this scheme, arising from dark radiation, rare $K$ decays, dark matter self-interactions, and gravity waves. In section \ref{sec:increasingv'} we explore the possibility of increasing $v'$ so that the freeze-out abundance of mirror electrons is too high, and subsequently reduced by entropy production from mirror neutrino decay. In the appendix, we present a model where the axion quality problem is solved such that the PQ symmetry is an accidental consequence of anomaly-free discrete symmetries.

\section{The Mirror Theory with Exact Parity and an Axion}
\label{sec:theory}
Motivated by mirror dark matter and the axion quality problem, we study a theory containing the Standard Model (SM), its $Z_2$ symmetric mirror (SM$'$), and a single axion field coupled to both sectors. In this section, we discuss  the Lagrangrian of the two sectors and determine the spectrum of particles.

\subsection{Lagrangian}
 The $Z_2$ symmetry maps the SM gauge group and particles into their $Z_2$ mirrors
\begin{align}
\label{eq:Paction}
 SU(3) \times SU(2) \times U(1) \; \; \leftrightarrow \; \; & SU(3)' \times SU(2)' \times U(1)' \nonumber \\
 q, \bar{u}, \bar{d}, \ell, \bar{e} \; \; \leftrightarrow \; \; & q', \bar{u}', \bar{d}', \ell', \bar{e} \nonumber \\
 H \; \; \leftrightarrow \; \; & H' \nonumber \\
 F^{\mu \nu}, W^{\mu \nu}, G^{\mu \nu} \; \; \leftrightarrow \; \; & F'^{\mu \nu}, W'^{\mu \nu}, G'^{\mu \nu} \nonumber \\
 \Psi, \bar{\Psi} \; \; \leftrightarrow \; \; & \Psi', \bar{\Psi}' \nonumber \\ 
  P \; \; \leftrightarrow \; \; & P  \, , 
\end{align}
where  primes indicate mirror fields, and matter is described by 2-component, left-handed, Weyl fields. $\Psi$ and its mirror, $\Psi'$, are heavy quarks charged under the PQ symmetry while $P = \frac{1}{\sqrt{2}}(s + f_a)e^{i a/f_a}$ is the PQ breaking field that contains the axion, $a$, as its angular mode and the saxion, $s$, as its radial mode.

The Lagrangian of the theory is given by
\begin{align}
    \mathcal{L} = \mathcal{L}_{\rm SM} + \mathcal{L}_{\rm SM'} +\mathcal{L}_{\rm Portal} + \mathcal{L}_{\rm PQ},
\end{align}
where $\mathcal{L}_{\rm SM}$ is the SM Lagrangrian up to dimension-5 containing Yukawa interactions, the canonical field kinetic energies, and the Weinberg neutrino operator $\ell \ell HH$. Similarly, $\mathcal{L}_{\rm SM'}$ is the associated mirror Lagrangian which is related to $\mathcal{L}_{\rm SM}$ by the mapping given in \eqref{eq:Paction}. 
The portal, $\mathcal{L}_{\rm Portal}$, contains the $Z_2$ symmetric operators that mix the SM and mirror sectors
\begin{align}
    \label{eq:Lportal}
\mathcal{L_{\rm Portal}} = \frac{\epsilon}{2} B_{\mu \nu}B'^{\mu \nu} -\lambda' |H|^2|H'|^2  + \left(\frac{\xi_{ij}}{M_D} \ell'^i H \; \ell^j H' +  \rm{h.c.} \right),
\end{align}
which we call the kinetic mixing portal, Higgs portal, and neutrino portal, respectively.

Last, the Lagrangian for the PQ breaking field and the PQ quarks is
\begin{align}
    \label{eq:LPQ}
    \mathcal{L}_{\rm PQ} = \xi (\bar{\Psi} \Psi + \bar{\Psi}' \Psi') P &+ \text{h.c.} + \lambda(|P|^2 - f_a^2/2)^2 
    + |{D}^\mu P|^2 
    +i\Psi^\dag \not\!{\partial} \Psi + \text{h.c.}
\end{align}
In Appendix~\ref{sec:accidental}, we present a model where an exact, extra $Z_{2n+1}$ symmetry ensures the quality of the PQ symmetry.

In the remainder of this work, we consider the effective theory below the scale $\langle P \rangle = f_a/\sqrt{2}$ which, after integrating out $\Psi$, generates the axion interactions 
\begin{align}
    \label{eq:Laxion}
    \mathcal{L}_a= \frac{1}{32\pi^2}\frac{a}{f_a} \left( g_3^2 \, G_{\mu \nu} \tilde{G}^{\mu \nu} + g_3'^2 \, G'_{\mu \nu} \tilde{G'}^{\mu \nu} \right) 
    + \frac{1}{32\pi^2} \frac{E}{N} \frac{a}{f_a} \left( e^2 \, F_{\mu \nu} \tilde{F}^{\mu \nu} + e'^2 \, F'_{\mu \nu} \tilde{F'}^{\mu \nu} \right)\, .
\end{align}
Note that above the mirror symmetry breaking scale $\langle H' \rangle = v' \gg v$, the $Z_2$ symmetry enforces $g_3 = g_3'$ and $e = e'$. Below this scale, the renormalization group running of the two sectors differs so that $g_3'$ and $e'$ can diverge from their SM counterparts.

\subsection{Breaking of Mirror and SM Electroweak Symmetries}

If the $Z_2$ symmetry of (\ref{eq:Paction}) is exact, the potential for $H$ and $H'$ is 
\begin{align}
    \label{eq:VH}
    V_{H,H'} \; = \; V_{\rm{SM}}(H) + V_{\rm{SM'}}(H') + \lambda' |H|^2 |H'|^2.
\end{align}
The vacuum, determined at loop level by the Higgs Parity mechanism \cite{Hall:2018let}, has one vev much larger than the other. Defining $H'$ to have the larger vev, $v'$, the low-energy effective potential of $H$ at tree-level is
\begin{align}
    \label{eq:V_LE}
    V_{\rm L.E.} = - 2\lambda v'^2\left(1 - \frac{\lambda'}{2\lambda}\right) |H|^2 + \lambda \left(1 - \frac{{\lambda'}^2}{4 \lambda^2}\right) |H|^4 \, .
\end{align}
Identifying the quadratic term with the SM Higgs mass squared, $|\lambda' - 2 \lambda |\ll 1$ is required to give $v \ll v'$. Eq.~\eqref{eq:V_LE}, then implies the quartic term of the Higgs potential is nearly zero at the scale $v'$ as it is proportional to $\lambda'- 2 \lambda$~\cite{Hall:2018let}. 

In the SM, the renormalization of the Higgs quartic is dominated by quantum corrections from the top quark which causes $\lambda$ to decrease from low to high energies and eventually become zero. From a perspective of running the Higgs quartic from low to high energies, we can thus identify $v'$ with the energy scale $\mu$ at which the Higgs quartic vanishes $\lambda(\mu = v') \simeq 0$. In the SM and in a pure mirror model, where the only particle content is the SM and its $Z_2$ mirror, the scale at which $\lambda = 0$ occurs is at $\mu \approx 10^{12}$ GeV for $m_{\rm top}$ and $\alpha_s$ at their central experimental values \cite{Buttazzo:2013uya,Workman:2022ynf,Dunsky:2019upk}. Furthermore,  $\mu$ is above $10^9 (10^{10})$ GeV at $3\sigma (2 \sigma)$.

However, as discussed in Sec.~\ref{sec:epDM}, $v' > 10^8$ GeV is problematic from a cosmological perspective because freeze-out of the stable mirror electron overproduces dark matter. One possibility is to dilute the mirror electrons by large entropy generation, for example from mirror neutrino decay \cite{Dunsky:2019upk}. Another possibility is to include particle content beyond the SM with interactions that makes $\lambda$ run faster so that it vanishes at the scale $\mu \approx 10^8$ GeV. In the mirror model, with an axion having $f_a < v'$ as considered in this paper, the PQ quarks $\Psi$ of Eq.~\eqref{eq:LPQ} are present in the effective theory below $v'$. Furthermore, they have Yukawa interactions with $H$ if their gauge charges are the same as one of the SM quark species, $q, \bar{u}$ or $\bar{d}$. For example, choosing the PQ quark to be an SU(2) singlet with hypercharge $1/3$, and calling it $D$, allows the Yukawa interaction
\begin{align}
    \label{eq:LYD}
    \mathcal{L}_D = y_D (q \bar{D} H + q' \bar{D'} H') \, .
\end{align}
Below $v'$, the first operator of Eq.~\eqref{eq:LYD} generates a quantum correction to the beta function of $\lambda \propto - y_D^4$. Reducing the scale at which $\lambda = 0$ from $10^{12}$ GeV to $10^{8}$ GeV can easily be accomplished for $y_D \sim \mathcal{O}(1)$.  Note that $\bar{D}$ is defined as the linear combination of anti-down type quarks that couples to $P$ via the coupling $\xi$ of (\ref{eq:LPQ}). Dark radiation bounds as discussed in Sec. \ref{sec:neff} constrain $f_a$ so that the values of $f_a$ of interest to us are of order 30 TeV. Consequently, with $\xi \sim \mathcal{O}(1)$ the mass of $D$ is also of order 30 TeV. 

On the other hand, the mass of $D'$ is complicated by the large vev of $H'$ which strongly mixes $D'$ with the down-type component of $q'$, $d'$, which may be the mirror down, strange, or bottom quark. The associated mass matrix is
\begin{align}
\label{eq:DpMassMatrix}
M_{d' D'} = 
    \begin{pmatrix}
        d' && D'
    \end{pmatrix}
    \begin{pmatrix}
        y_d v' && y_D v' \\
        0   &&  \xi f_a
    \end{pmatrix}
    \begin{pmatrix}
        \bar{d'} \\
        \bar{D'}
    \end{pmatrix}
    + \text{h.c.}
\end{align}
Eq.~\eqref{eq:DpMassMatrix} possesses a heavy eigenvalue of mass $m_{D, \rm heavy} \sim y_D v' \sim v'$ and a light eigenvalue of mass $m_{D, \rm light} \sim \xi f_a y_d/y_D \sim y_d f_a$. As long as $y_d$ is the bottom quark Yukawa coupling, then $m_{D, \rm light}$ is greater than the mirror QCD scale (see Eq.~\eqref{eq:mirrorQCDscale}) and the $SU(3)'$ sector remains an effective zero-flavor Yang-Mills theory. In this scenario, the $SU(3)'$ phase transition is first order and a strong gravitational wave signal from bubble nucleation is generated as discussed in Sec. \ref{sec:GW}.

Another option for obtaining $v' = 10^8$ GeV, needed for $e'$ dark matter from freeze-out, is to add a soft $Z_2$-breaking term to the Higgs potential of (\ref{eq:V_LE}): $\delta^2 (|H|^2 - |H'|^2)$.  In this case, the PQ quarks need not couple to the Higgs; indeed, they may be neutral under the electroweak gauge group.  This is important as the dark radiation signal, discussed in Sec. \ref{sec:neff}, is very sensitive to the electromagnetic anomaly of the PQ symmetry.

\section{Particle Spectra and Lifetimes}
\label{sec:spectra}
\begin{figure}[tb]
    \centering
    \includegraphics[width=.75\textwidth]{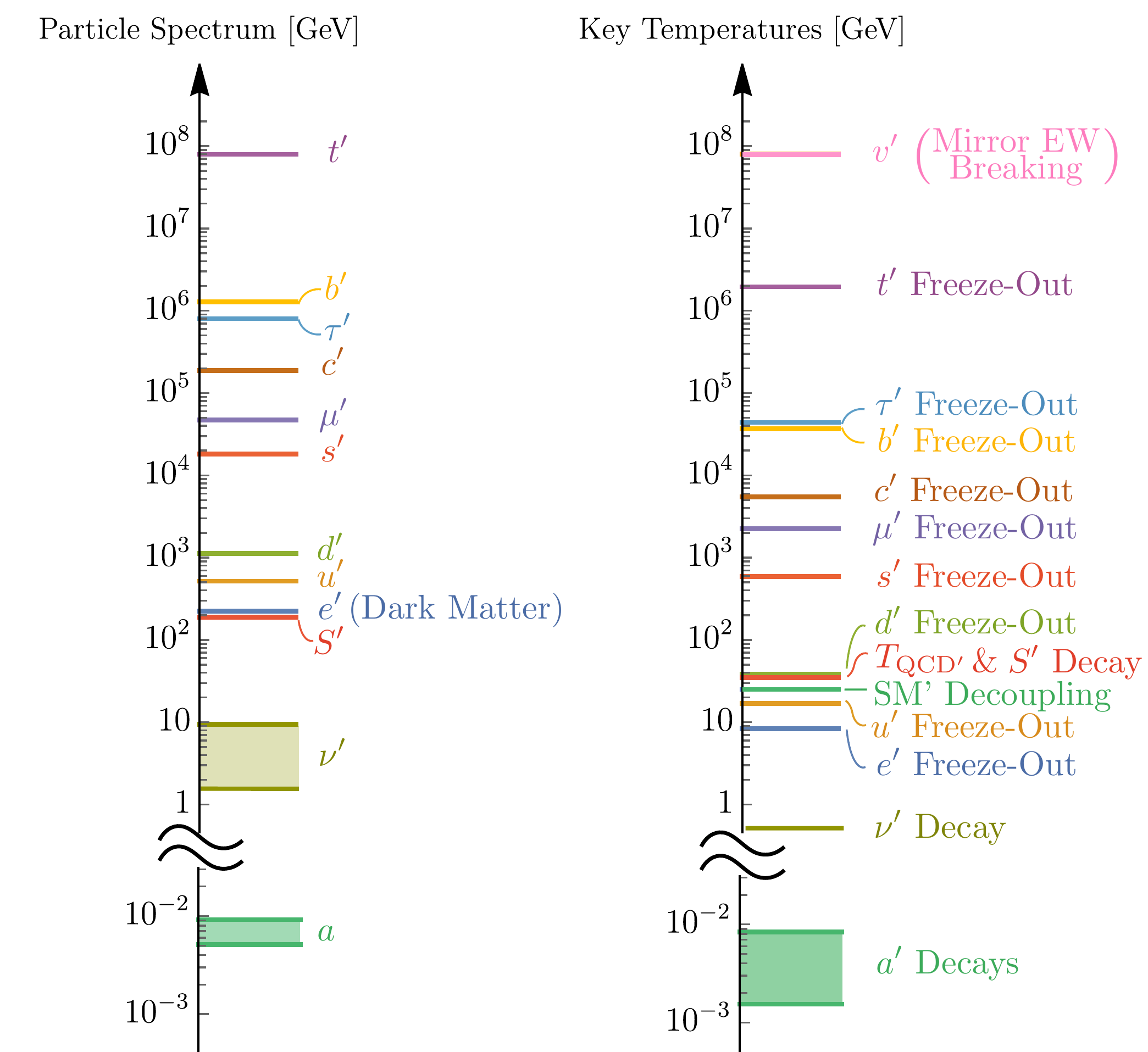} 
    \caption{Left panel: spectrum of mirror quarks and leptons for $v' = 8 \times 10^7$ GeV, which yields the observed dark matter abundance in $e'$ from freeze-out. The mass of the lightest mirror glueball, $S'$, is also shown, as is the relevant range of axion masses. Right panel: important temperatures in the thermal history of the universe. The mirror electroweak scale, $v'$, is also shown.}
    \label{fig:spectrum}
\end{figure}
\subsection{Mirror Fermions}
Below the mirror symmetry breaking scale, $v'$, charged mirror fermions acquire a mass $m' = y v'$ where $y$, their Yukawa coupling with $H'$, is identical to the Yukawa coupling of their SM counterparts, up to renormalization differences below $v'$. Consequently, charged mirror fermions, such as the mirror electron, $e'$, are heavier than their SM counterparts by approximately the ratio of electroweak vacuum expectation values, $v'/v \gg 1$. The spectrum of mirror particles, including the renormalization of the Yukawa couplings below $v'$, is shown in Fig.~\ref{fig:spectrum}. 

Note that all mirror fermions are unstable except for $e'$ and $u'$ which cannot decay since they are the lightest particles charged under the unbroken $U(1)_{\rm EM}'$ and $SU(3)'$ symmetries, respectively. 
Since $e'$ is stable and has suppressed interactions with SM particles, its relic cosmological abundance may provide the dark matter. As discussed in Sec. \ref{sec:epDM}, thermal freeze-out of $e'$ leads to the observed dark matter abundance if $v' \simeq 8 \times 10^7$ GeV, setting our normalization for $v'$ in all future equations. The mass of $e'$ is
\begin{align}
    m_e' = y_e v' \simeq 225\; {\rm GeV} \left(\frac{v'}{8 \times 10^7 \, {\rm GeV}}\right).
\end{align}
Note that while $u'$ is also stable and a potential dark matter candidate, its relic abundance is always subdominant to that of $e'$. This is because below the mirror QCD phase transition, mirror hadrons composed of $s'$, $u'$ and $d'$ efficiently annihilate into $\gamma'$ or decay into the hadron $u' u' u'$ whose abundance is small compared to $e'$ \cite{Dunsky:2019upk}.

\subsection{Axion}
At energy scales above $v'$ the $Z_2$ symmetry ensures that the QCD and QCD$'$ gauge couplings are equal, $g_3 = g_3'$. However, below $v'$ the $Z_2$ symmetry is broken, and the large mass hierarchy between the Standard and mirror fermions significantly changes the renormalization flow of $g_3'$ away from $g_3$. As the heavier mirror fermions decouple in the IR, $g_3'$ becomes larger than $g_3$, leading to a higher confinement scale in the $SU(3)'$ sector. The value of $g_3$ and the renormalization scale $\mu = v' = 8 \times 10^7$ GeV sets the initial value of $g_3'$ at the same scale. We then run the 2-loop beta function for $g_3'$ \cite{Caswell:1974gg,Jones:1974mm,Machacek:1983tz} to low energies, taking into account the mirror quark thresholds. We define $\Lambda_{\rm QCD'}$ as the scale where $g_3'(\mu)$ that is defined in the $\overline{MS}$ scheme and computed at two-loop level diverges.
Around $v' \sim 10^8$ GeV, it is given by
\begin{align}
    \label{eq:mirrorQCDscale}
    \Lambda_{\rm QCD'} \simeq 27 {\, \rm GeV} \left(\frac{v'}{8 \times 10^{7} \, \rm GeV} \right)^{4/11} \, .
\end{align}
Consequently, the \textit{single} axion in the theory, which couples to both Standard Model and mirror sectors, acquires a large mass from mirror QCD instantons,
\begin{align}
    \label{eq:ma}
    m_a = 0.6 \frac{\Lambda_{\rm QCD'}^2}{f_a} \simeq 9 \,{\rm MeV} \left(\frac{v'}{8 \times 10^{7} \, \rm GeV}\right)^{8/11} \left(\frac{f_a}{5 \times 10^{4} \, \rm GeV}\right)^{-1} \, ,
\end{align}
where we have used the lattice calculations for the topological susceptibility of $N_f = 0$ flavor $SU(3)$ and the relation between $\Lambda_{\rm QCD'}$ and the Sommer scale~\cite{Durr:2006ky,FlavourLatticeAveragingGroupFLAG:2021npn} to relate the mass of the axion with $\Lambda_{\rm QCD'}$.
The axion mass is thus much larger than the standard QCD axion mass $m_{a,0} \sim m_\pi f_\pi/f_a$. Moreover, the mapping given in \eqref{eq:Paction}  ensures the phases of the potential generated by $SU(3)$ and $SU(3)'$ dynamics are aligned so that the strong CP problem is still solved by the axion, despite its non-standard mass. The heavier axion mass relaxes the PQ quality problem that plagues the standard QCD axion \cite{Kamionkowski:1992mf,Barr:1992qq,Ghigna:1992iv,Holman:1992us}. 

For $m_a \ll m_{\pi^0} \simeq 135$ MeV, the lifetime of the axion is set by the $a \rightarrow 2\gamma$ and $a \rightarrow 2\gamma'$ decay channels. The total decay rate is
\begin{align}
    \label{eq:axionDecayRate}
    \Gamma_{a} = \frac{g_{\gamma}^2 + g_{\gamma'}^2}{64\pi}\frac{m_a^3}{f_a^2} \simeq \left(1.5 \times 10^{-1} \, \rm s \right)^{-1}\left(\frac{m_a}{9 \, \rm MeV}\right)^3 \left(\frac{f_a}{5 \times 10^4 \, \rm GeV} \right)^{-2} \, ,
\end{align}
where the axion-photon and dark photon couplings are $g_\gamma \simeq (E/N - 5/3 - \mathcal{F_\theta}(m_a))\alpha/2\pi$ and $g_{\gamma'} \simeq (E/N)\alpha'/2\pi$, respectively. Here, $E/N$ is the ratio of the charge to color anomaly of the PQ quarks, $5/3$ comes from the axial rotation that removes the axion-gluon coupling,  $\mathcal{F}_\theta$ from axion-meson mixing \cite{Aloni:2018vki}, and $\alpha \, (\alpha')$ the (mirror) fine-structure constant. $E/N = 2/3$ when the PQ quarks have the same charges as down-quarks and $8/3$ if they have the same charges as up-quarks or in a complete $SU(5)$ multiplet. For $m_a \ll m_\pi$, as occurs for the axion in most of our parameter space, $(5/3 + \mathcal{F}_\theta) \simeq 2.03$. For Eq.~\eqref{eq:axionDecayRate} and the remainder of this paper, we take $\alpha' = \alpha$ due to the small renormalization differences between the SM and SM$'$ electromagnetic sectors.

\subsection{Mirror Glueballs}
Mirror glueballs materialize after the QCD$'$ phase transition. Self scattering quickly leaves the glueball gas in a state dominantly composed of the lightest mirror glueball, $S'$, with a mass \cite{Durr:2006ky}
\begin{align}
    m_{S'} \simeq 6.8 \; \Lambda_{\rm QCD'} \simeq 185 \, {\rm GeV} \left(\frac{v'}{8 \times 10^7 \, {\rm GeV}} \right)^{4/11}. 
\end{align}
The mirror glueball can decay into two Higgses, two mirror photons, or two axions; it can also self-scatter with itself to reduce its abundance. Each rate is strongly dependent on $v'$, but for $v' = 10^8$ GeV, the rates increasingly dominate in the aforementioned order. The decay rate into electroweak gauge bosons is given by \cite{Dunsky:2019upk}
\begin{align}
     \Gamma_{S' \rightarrow W^+W^-/ZZ} \simeq 
    \frac{1}{8\pi} \left(\frac{2.7}{16\pi^2}\right)^2 \frac{m_{S'}^5}{v'^4} \simeq \left(11 \, {\rm s}\right)^{-1} \left(\frac{v'}{8 \times 10^7 \, \rm GeV}\right)^{-\frac{24}{11}} \, .
\end{align}
Here we ignored the phase-space suppression and the actual decay rate is even smaller.
The decay rate into $2\gamma'$ is given by
\begin{align}
    \Gamma_{S' \rightarrow \gamma' \gamma'} \simeq \frac{1}{16\pi}\left(\frac{2.7 \alpha}{270 \pi} \right)^2 \frac{m_{S'}^9}{m_{u'}^8} \simeq \left(1.1 \times 10^{-12} \, {\rm s}\right)^{-1} \left(\frac{v'}{8 \times 10^7 \, \rm GeV}\right)^{-\frac{52}{11}} \, .
\end{align}
The decay rate into $2a$ is given by
\begin{align}
    \Gamma_{S' \rightarrow a a} \approx \frac{1}{8 \pi} \frac{m_{S'}^5}{f_a^4} \simeq \left(4.8 \times 10^{-16} \, {\rm s}\right)^{-1}  \left(\frac{f_a}{5 \times 10^4 \, \rm GeV}\right)^{-4} \left(\frac{v'}{8 \times 10^7 \, \rm GeV}\right)^{\frac{20}{11}} \, .
\end{align}
Last, the annihilation of two lightest glueballs (CP-even states) into an axion and the lightest CP-odd glueball is of similar importance in reducing the glueball abundance, and has a rate
\begin{align}
    \label{eq:glueballscatteringrate}
    \Gamma_{S' S' \rightarrow S' a} &\sim \frac{\left(m_{S'} T\right)^{\frac{3}{2}}}{f_a^2} e^{-\frac{m_S'}{T}} \nonumber \\
    &\approx  \left(7.1 \times10^{-22} \, {\rm s}\right)^{-1}  \left(\frac{f_a}{5 \times 10^4 \, \rm GeV}\right)^{-2} \left(\frac{v'}{8 \times 10^7 \, \rm GeV}\right)^{\frac{12}{11}} \left(\frac{T}{m_S'}\right)^{\frac{3}{2}}e^{-\frac{m_S'}{T}} .
\end{align}
\subsection{Mirror Neutrinos and Neutrinos}

A successful theory of $e'$ dark matter from freeze-out has important implications for neutrino masses because late decays of mirror neutrinos can dilute the $e'$ and can affect nucleosynthsis. Parity implies that the SM and SM$'$ Weinberg operators give correlated masses for the SM and mirror neutrinos. In addition, the neutrino portal operator of (\ref{eq:Lportal}) gives a neutrino Yukawa coupling.  Hence, below $v'$ the relevant EFT for neutrino masses is
 \begin{align}
  \label{eq:Lnunu'}
     {\cal L}_{\nu,\nu'} = \frac{1}{2} m_{\nu'_i} \left( \nu'_i \nu'_i
     + \ell_i \ell_i \, \frac{HH}{v'^2}\right) + \nu'_i \, y_{ij} \, \ell_j \, H + {\rm h.c.}, \qquad y_{ij} = \frac{v'}{M_D} \xi_{ij}.
\end{align}
The light neutrino mass matrix is therefore
 \begin{align}
 \label{eq:mnuij}
     m_{\nu_{ij}} = \frac{v^2}{v'^2} \, m_{\nu'_i} \; \delta_{ij} - y^T_{ik}\, \frac{1}{m_{\nu'_k}} \, y_{kj} \; v^2.
 \end{align}
We call the first term the ``direct" contribution and the second the ``seesaw" contribution.

It is useful to study the direct and seesaw contributions to the light neutrino mass matrix  from each $\nu'_i$. Each $\nu'_i$ couples to a single combination of $\ell_j$, which we call $\tilde{\ell}_i$, so that the Yukawa coupling of $\nu'_i$ can be written as
 \begin{align}
 \label{eq:Lyi}
     {\cal L}_{y_i} = y_i \; \nu'_i \, \tilde{\ell_i} \, H + {\rm h.c.}, \qquad \tilde{\ell_i} = y_{ij} \, \ell_j/y_i, \qquad y_i^2 \equiv \sum_j |y_{ij}|^2.
\end{align}
Thus, each $\nu'_i$ gives mass contributions to two different states, a direct one for $\nu_i$ and a seesaw one for $\tilde{\nu}_i$. If large leptonic mixing angles arise from the neutrino sector, these two states are expected to be very different, although generically they are not orthogonal. Consequently, the Lagrangian for the light neutrino masses can be written as a sum of three such terms, one from each $\nu'_i$
 \begin{align}
 \label{eq:Lnu}
     {\cal L}_\nu =\frac{1}{2} \sum_i \left( m_{\nu_i}^{\rm dir} \; \nu_i \nu_i - m_{\nu_i}^{\rm ss} \; \tilde{\nu}_i \tilde{\nu}_i \right) + {\rm h.c.} =\frac{1}{2} \sum_i \left(\frac{v^2}{v'^2} \; m_{\nu'_i} \;  \nu_i \nu_i
     - \; \frac{y_i^2 v^2}{m_{\nu'_i}} \; \tilde{\nu}_i \tilde{\nu}_i \right) + {\rm h.c.}.
\end{align}
The resulting neutrino mass matrix therefore has 6 terms and depends on $(v', m_{\nu'_i}, y_i)$ as well as the mixing angles and phases in $y_{ij}/y_i$.

\begin{figure}[tb]
    \centering
    \includegraphics[width=.75\textwidth]{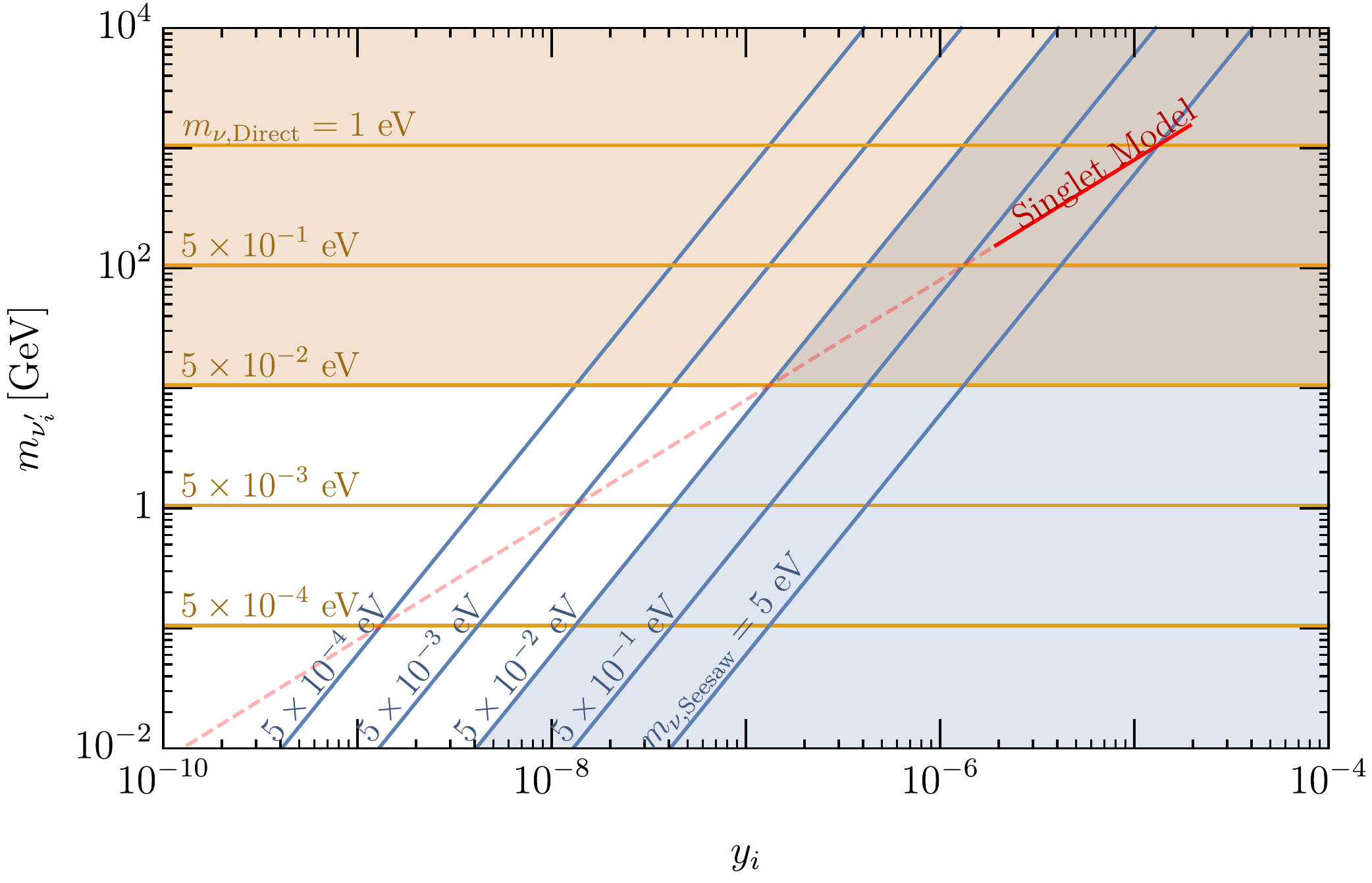} 
    \caption{Contours for the direct (orange) and seesaw (blue) contributions to the light neutrino mass matrix, from the heavy mirror neutrino mass eigenstate $\nu'_i$, as defined in Eq.~(\ref{eq:Lnu}), with $v' = 8 \times 10^7$ GeV. The orange (blue) shaded regions give $m_{\nu_i}^{\rm dir} (m_{\nu_i}^{\rm ss})$ larger than 0.05 eV, and in many theories require unnatural cancellations. The red line has $m_{\nu_i}^{\rm dir} = m_{\nu_i}^{\rm ss}$, giving massless light neutrinos at tree level. The Singlet Model is constrained to live on this line for all three $\nu'_i$. In this theory, radiative neutrino masses of 0.05 eV can result along the solid part of the red line.}
    \label{fig:mnuDilutionPlot}
\end{figure}

The orange and blue contours of Fig.~\ref{fig:mnuDilutionPlot} show the direct and seesaw contributions from one $\nu'_i$ in the $(y_i,m_{\nu'_i})$ plane. For the direct contribution, the scale $v'$ of parity breaking is taken to be $8 \times 10^7$ GeV, as required for $e'$ to account for the observed dark matter via freezeout. The mass of $\nu'_i$ is directly proportional to the direct neutrino mass contribution
\begin{align}
    m_{\nu_i'} = 11 \, {\rm GeV} \left( \frac{m_{\nu_i}^{\rm dir}}{0.05 \, \rm eV}\right)\left(\frac{v'}{8 \times 10^7 \, {\rm GeV}}\right)^2 \, .
    \label{eq:numassWeinberg}
\end{align}
In the orange (blue) shaded regions the direct (seesaw) contribution is larger than 0.05 eV, so a realistic spectrum requires a cancellation among the 6 contributions. Deeper into the shaded region the required cancellation becomes stronger. 
Deep in the unshaded region, both direct and seesaw contributions are too small to affect current data. There are three copies of this figure, one for each $\nu'_i$. Assuming no strong cancellations, at least 2 of the 6 contributions must be close to the edge of the corresponding shaded region to account for the atmospheric and solar oscillations. Two relevant contributions could come from a single $\nu'_i$ if the values of $(y_i,m_{\nu'_i})$ lie close to to the right-hand vertex of the unshaded triangle, where the 0.05 eV contours of the direct and seesaw contributions intersect.  This intersection point gives the maximum value of $y_i$ possible without cancellations, $y_i \lesssim m_{\nu'_i}/v'$, and similarly the maximum value of $|y_{ij}|$ for any $j$ 
\begin{align}
    \label{yijbound}
    |y_{ij}| \lesssim \frac{m_{\nu'_i}}{v'} \approx 1 \times 10^{-7} \left(\frac{m_\nu}{0.05 \, \rm eV}\right) \left(\frac{v'}{8 \times 10^7 \, \rm GeV} \right) \, .
\end{align}

Fig.~\ref{fig:mnuDilutionPlot} shows that the $\nu'_i$ have masses far below $v'$, so that decays via virtual $W'$ are negligible and $\nu'_i$ have long lifetimes.  Decays proceed dominantly through the Yukawa coupling $y_i$ of \eqref{eq:Lyi}, which induces a small mixing angle between $\nu'_i$ and $\tilde{\nu}_i$ of
\begin{align}
    \theta_i \simeq  \frac{y_i v}{m_{\nu'_i}}. 
\end{align}
For $m_{\nu'_i} \lsim m_{W} + m_{e}$, the dominant decay modes are the three-body beta decays $\nu'_i \rightarrow \tilde{e}_i \,  u_j \, \bar{d}_j, \, \tilde{e}_i \,  \nu_j \, \bar{e}_j$, with $j$ running over two generations of quarks and three generations of leptons, giving 
\begin{align}
    \label{eq:nuPlifetime3}
    \Gamma_{\nu'_i} \; \simeq \; 9\frac{1}{192 \pi^3}\; \theta_i^2 \; \frac{m_{\nu'_i}^5}{8v^4} \, .
\end{align}
Similarly, there are $Z$-mediated decay modes. For $m_{\nu'_i} \gtrsim m_{W} + m_{e}$, the two-body decay $\nu'_i \rightarrow \tilde{\ell}_i H$ becomes kinematically allowed and dominates, with a decay rate
\begin{align}
    \label{eq:nuPlifetime2}
    \Gamma_{\nu'_i} \; \simeq \; 
    \frac{1}{8\pi} \,  y_i^2 \; m_{\nu'_i}
     \, .
\end{align}

\subsubsection{Neutrino masses constrained by lepton symmetries}
\label{sec:diracnu}

The general form of neutrino masses, (\ref{eq:Lnunu'}), has two important limits, one where $m_{\nu'_i}=0$ and one where $y_{ij}=0$. This changes the physics drastically, and Fig.~\ref{fig:mnuDilutionPlot} is no longer relevant. 

Imposing $B-L$ number,
with opposite signs in the SM and mirror sectors, sets $m_{\nu'_i}=0$ so that the observed neutrinos are Dirac. The heavy, unstable $\nu'$ states do not exist, changing the cosmological evolution of the universe. Unlike the charged fermions, where the right-handed states are part of the SM sector, neutrinos have their right-handed components  coming from the mirror sector, allowing an alternative view of why the neutrinos are so light. The observed neutrino masses result from Yukawa couplings $y_{ij}$ of order $10^{-12}$.  However, unlike for charged fermions, these Yukawa interactions arise from dimension 5 operators above $v'$, shown in (\ref{eq:Lportal}). With $v' = 10^8$ GeV and $M_D$ the Planck scale, this requires $\xi_{ij}$ of order $10^{-2}$, comparable to the largest Yukawa coupling in the charged lepton sector. 

Alternatively, separate lepton parities can be imposed in the SM and mirror sectors, removing the neutrino portal operator of (\ref{eq:Lportal}) and forcing $y_{ij}=0$.  The neutrino sector then conserves individual lepton parities for each of the three generations, and the SM and mirror neutrino masses are related by $m_{\nu_i} = (v^2/v'^2)m_{\nu'_i}$. Lepton mixing angles arise because in this basis the Yukawa coupling matrix for the charged leptons is not diagonal. For $v' = 10^8$ GeV, the lightest mirror neutrino, $\nu'_1$, is stable as it is the lightest mirror fermion. The two heavier mirror neutrinos decay via a loop diagram involving a virtual $W'$, $\nu'_{2,3} \rightarrow \nu'_1 + \gamma'$. The decay rate is highly suppressed, giving lifetimes of order $10^9$ sec.

\subsubsection{The singlet model for neutrino masses}
\label{sec:singlet}
In many theories, significant cancellations between different contributions to the light neutrino mass matrix can only arise from fine tuning.  In these theories the shaded regions of Fig.~\ref{fig:mnuDilutionPlot} are fine-tuned, and (\ref{yijbound}) are naturalness constraints. However, cancellations cannot be avoided in one of the simplest UV completions of the dimension 5 operators for neutrino masses~\cite{Hall:2019qwx}, which we call the Singlet Model.

Consider a theory where the dimension 5 operators for neutrino masses are generated by the exchange of three gauge-singlet Weyl fermions $S_i$, that are parity even: $S_i \leftrightarrow S_i$, via the interactions
\begin{align}
    \label{eq:LSi}
    {\cal L}(S_i) = S_i \, x_{ij} \,  (\ell_j H + \ell'_j H') +  
    \frac{1}{2} M_{S_i} \, S_i S_i + {\rm h.c.}
\end{align}
Introducing a hatted basis, the Yukawa interactions of $S_i$ can be written as 
 \begin{align}
 \label{eq:Lxi}
     {\cal L}_{x_i} = x_i \; S_i \, (\hat{\ell}_i \, H +\hat{\ell'}_i \, H') + {\rm h.c.}, \qquad \hat{\ell_i} = x_{ij} \, \ell_j/x_i, \qquad x_i^2 \equiv \sum_j |x_{ij}|^2,
\end{align}
in analogy with (\ref{eq:Lyi}).  The EFT below $M_{S_i}$ is
 \begin{align}
 \label{eq:LxiEFT}
     {\cal L}_{EFT}(S_i) \; = \; \frac{1}{2} \, \sum_i \frac{x_i^2}{M_{S_i}} \;  (\hat{\ell}_i \, H +\hat{\ell'}_i \, H')^2 + {\rm h.c.}.
\end{align}
Note that the $\hat{\ell}_i$ are not orthogonal.

Inserting the $H'$ vev, and comparing with Eqns. \eqref{eq:Lnunu'} and \eqref{eq:Lyi}, the singlet model leads to a correlation between $m_{\nu'_i} = x_i^2v'^2/M_{S_i}$ and $y_i = x_i^2v'/M_{S_i}$. Thus it gives $y_i = m_{\nu'_i}/v'$, leading to 
 \begin{align}
 \label{eq:dir=ss}
   m_{\nu_i}^{\rm dir} = m_{\nu_i}^{\rm ss} = \frac{y_i^2 v^2}{m_{\nu'_i}} = \frac{x_i^2 v^2}{M_{S_i}}  
\end{align}
giving the red line in Fig.~\ref{fig:mnuDilutionPlot}. 
The light neutrinos are massless at tree level  because each $S_i$ couples to only one combination of mirror neutrinos and neutrinos, $v' \hat{\nu}'_i + v\, \hat{\nu}_i$, leaving the orthogonal combination, $v' \hat{\nu}_i - v\, \hat{\nu}'_i$, without a mass term. This single coupling of $S_i$ leads to the combination $(\hat{\ell}_i \, H +\hat{\ell'}_i \, H')^2$ appearing in  (\ref{eq:LxiEFT}), implying that the coefficient of the Weinberg operators involving $\hat{\ell}_i\hat{\ell}_i$ and $\hat{\ell}'_i\hat{\ell}'_i$ are correlated with that of the portal operator involving $\hat{\ell}'_i\hat{\ell}_i$.  However, this correlation is lost under renormalization group scaling induced by one-loop electroweak radiative corrections. Hence the light neutrino mass arising from the interaction of $S_i$ is estimated to be
 \begin{align}
 \label{eq:1loopmass}
   {\cal L}_\nu(S_i) \; \simeq \; \frac{1}{2} \; \frac{y_i^2 \, v^2}{m_{\nu'_i}} \; \frac{g_2^2}{16 \pi^2} (2L_+ + L_-) \; \hat{\nu}_i \hat{\nu}_i,  \qquad L_+ = \ln \frac{M_{S_i}}{v'}, \; L_-= \ln \frac{v'}{m_{\nu'_i}}.
\end{align}
The two logs correspond to running in the EFTs above and below $v'$. Since $L_-$ is always a large log, the loop factor is expected to be in the range of 3-30, giving the extent of the solid section of the red line in Fig.~\ref{fig:mnuDilutionPlot}.

\section{Mirror Cosmology}
\label{sec:cosmology}
After inflation, the maximum temperature of the universe is taken to be less than $v'$, so that the spontaneous breaking of parity does not lead to unacceptable domain wall densities, but greater than the freeze-out temperature of mirror electrons, allowing a reliable computation of the relic $e'$ density.  

\subsection{Portals and Decoupling Temperatures}
\label{subsec:TFO}
Thermalization of the SM and SM$'$ sectors can be achieved through the Higgs, kinetic mixing,  neutrino, and heavy axion portals arising from the operators in Eqns.~\eqref{eq:Lportal} and \eqref{eq:Laxion}. We shall assume $\epsilon$ is negligibly small so that kinetic mixing is unimportant from here on.%
\footnote{The leading non-zero radiative correction to $\epsilon$ arises at the five-loop level from the interactions between the axion field $P$, the PQ quarks $\Psi, \Psi'$, and Higgses. Its value is of order $\epsilon_{\rm rad} \sim 1/(16\pi^2)^5 |\xi|^4 |y_D|^4 N_c^2$.  For $|y_\Psi|^4 |y_D|^4 \lesssim 0.1$, Rutherford scattering of $e'$ DM with nuclei in the LZ detector is below detection threshold for 15 ton-years of exposure \cite{Dunsky:2018mqs}.}

The Higgs and neutrino portals give rise to similar decoupling temperatures which increase with $v'$ and are independent of $f_a$. On the other hand, the axion portal gives rise to a decoupling temperature which increases with $f_a$ but is independent of $v'$ for temperatures above the QCD$'$ phase transition:
\begin{align}
    T_{\rm dec} \simeq 
    \label{eq:Tdec}
    \begin{dcases}
        5 \times 10^4 \, {\rm GeV} \left(\frac{v'}{1 \times 10^8 \, {\rm GeV}}\right)^{4/3} \quad &\text{Higgs and Neutrino Portals} \,  \text{\cite{Dunsky:2018mqs}} \\
        5 \times 10^4 \, {\rm GeV}  \left(\frac{f_a}{1 \times 10^9 \, {\rm GeV}}\right)^{2.35}  \quad &\text{Axion Portal~ \text{\cite{Salvio:2013iaa,DEramo:2021psx,DEramo:2021lgb}}} \\
    \end{dcases}
\end{align}
As discussed in Sec.~\ref{sec:epDM}, thermally producing $e'$ with the observed dark matter abundance requires $v' \approx 10^8$ GeV. From Eq.~\eqref{eq:Tdec}, we thus see that for $f_a \lesssim 10^9$ GeV, the axion portal dominates over the Higgs and Neutrino portals in keeping the two sectors in thermal equilibrium. Moreover, as discussed in Sec. \ref{sec:neff}, to ensure that the axion does not generate significant dark radiation around or after neutrino decoupling requires $f_a \lesssim 10^5$ GeV. Consequently, in our cosmology, the axion is the dominant portal responsible for thermal coupling between the two sectors.

Specifically, the axion keeps the two sectors in equilibrium both above and below the mirror QCD phase transition temperature,
\begin{align}
    \label{eq:TQCD'}
   T_{\rm QCD'} \; \simeq \; 1.26 \, \Lambda_{\rm QCD'} \; \simeq \; 34 {\, \rm GeV} \left(\frac{v'}{8\times 10^{7} \, \rm GeV} \right)^{4/11},
\end{align}
where we used the relation between $T_{\rm QCD'}$ and $\Lambda_{\rm QCD'}$ computed in~\cite{Borsanyi:2012ve}.
Above $T_{\rm QCD'}$,  $a \leftrightarrow g g$ and $a \leftrightarrow g'g'$ scatterings keep the two sectors in equilibrium. Below the QCD phase transition temperature, the two sectors are kept in thermal equilibrium through Primakoff scatterings, $\gamma e \leftrightarrow a e$  and $\gamma' e' \leftrightarrow a e'$, until the $e'$ number density drops exponentially at temperatures below $m_{e'}$.

The first interaction, that between axions and mirror gluons, can be calculated precisely from the standard axion-gluon scattering rate as computed in \cite{Salvio:2013iaa,DEramo:2021lgb} with the mapping $g_3 \leftrightarrow g_3'$,
\begin{align}
    \label{eq:agpRate}
    \Gamma_{ag' \leftarrow g'g'} \simeq \frac{16}{\pi} \left(\frac{g_3'^2}{32\pi^2} \right)^2 \frac{T^3}{f_a^2} \mathcal{F}_{g'} (g_3') \, .
\end{align}
Here, $\mathcal{F}_{g'}$ is a function of $g_3'$ and is numerically computed in \cite{Salvio:2013iaa,DEramo:2021lgb} but takes the form $\mathcal{F}_{g'} \approx 2 g_{3'}^2 \ln 1.5 /g_3'$ for $g_{3'}\lesssim 1$. 
\footnote{ Note that Eq.~\eqref{eq:agpRate} is strictly valid in the limit $m_a \ll T$, which is always the case for temperatures above $T_{\rm QCD'}$.}
When using Eq.~\eqref{eq:agpRate}, we use the temperature-dependent value of $g_3'$ at the renormalized energy scale $\mu = T$.

Similarly, the mirror Primakoff scattering rate is given by the Standard Model rate \cite{Bolz:2000fu,Cadamuro:2011fd} with the replacement of the electron number density with the mirror electron density, $n_{e} \rightarrow n_{e'}$, as given by
\begin{align}
    \label{eq:aDarkPrim}
    \Gamma_{a e' \leftrightarrow e' \gamma'} \simeq \alpha' \left(\frac{g_{\gamma'}}{f_a}\right)^2 \frac{\pi^2}{36 \zeta(3)} \left(\ln \frac{T^2}{m_\gamma'^2} + 0.82 \right) n_{e'} \, ,
\end{align}
where $g_{\gamma'} \simeq (E/N)\alpha'/2\pi$ and $m_\gamma'^2 \approx 4\pi \alpha' n_{e'}/m_{e'}$ is the mirror photon plasma mass. Because the renormalization in the electromagnetic gauge couplings are small, we take $\alpha' = \alpha$.

\begin{figure}[tb]
    \centering
    \includegraphics[width=.9\textwidth]{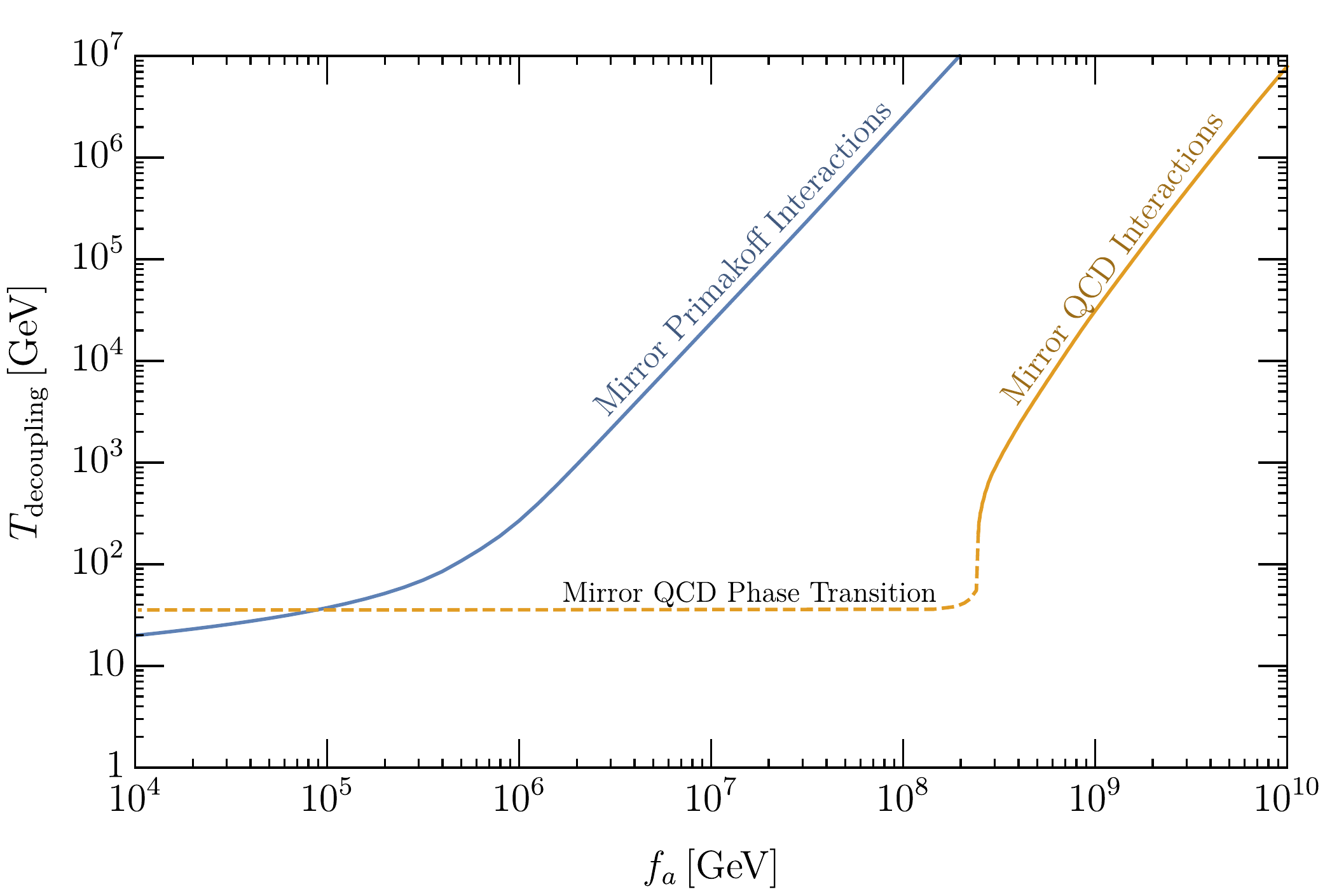} 
    \caption{Decoupling temperature of SM and mirror sectors from the axion portal as a function of $f_a$. The orange curve shows the decoupling temperature from axion and mirror gluon interactions which become ineffective around the mirror QCD phase transition at $T_{\rm QCD'} \approx 34$ GeV. The blue curve shows the decoupling temperature from axion and mirror electron scatterings (Primakoff) which become ineffective at temperatures below $m_e' \approx 225$ GeV . For $f_a \lesssim 10^5$ GeV, the decoupling temperature from either interaction are comparable and occurs around $20-30$ GeV.}
    \label{fig:axionDecoupling}
\end{figure}

We define the temperature when the SM and SM' sectors decouple by the temperature at which the rate $\Gamma_{a \leftrightarrow \rm{SM'}} = 3H$, where $H$ is Hubble and $\Gamma_{a \leftrightarrow \rm{SM}'}$ is the sum of the rates given in Eqns.~\eqref{eq:agpRate} and \eqref{eq:aDarkPrim}. The axion decoupling temperature from the mirror Standard Model is shown in Fig.~\ref{fig:axionDecoupling} as a function of $f_a$ for fixed $v' = 8 \times 10^7$ GeV ($\Lambda_{\rm QCD'} \approx 27$ GeV, $m_e' \simeq 225$ GeV), which is the required $v'$ to freeze-out mirror electrons as dark matter (see Sec. \ref{sec:epDM}). The solid orange curve shows the axion decoupling temperature from mirror gluon interactions, Eq.~\eqref{eq:agpRate}, which flatlines at $T_{\rm QCD'} \approx 34$ GeV when $g_{3'}$ becomes non-perturbative. Note that since $u'$ and $d'$ are much heavier than $\Lambda_{\rm QCD'}$, mirror pions are heavy and thus severely Boltzmann suppressed in number density immediately after the mirror QCD phase transition. Thus, we do not expect mirror quark bound states to significantly reduce the decoupling temperature below $T_{\rm QCD'}$ even for $f_a$ below $\sim 10^8$ GeV. Consequently, we extrapolate the decoupling temperature due to mirror gluons to lower $f_a$ by the dashed horizontal orange line. Similarly, the blue contour of Fig.~\ref{fig:axionDecoupling} shows the axion decoupling temperature from mirror Primakoff interactions. We see that for $f_a \lesssim 10^5$ GeV, the Primakoff interactions can dominate, keeping the two sectors in equilibrium to temperatures slightly below $\Lambda_{\rm QCD'}$. Since $f_a \gtrsim 10^5$ GeV produces long-lived axions that decay and generate too large $|\Delta N_{\rm eff}|$ as discussed in Sec. \ref{sec:neff}, we will generally be in the low $f_a$ region where Primakoff interactions set the decoupling temperature between the two sectors to $T_{\rm decouping} \approx 20-30$ GeV.

\subsection{$e'$ Dark Matter}
\label{sec:epDM}
The stability of $e'$ and $u'$ guarantees the existence of dark matter in the theory. The freeze-out chronology is similar to \cite{Dunsky:2019upk}, except for the existence of the axion, which maintains thermal equilibrium between the two sectors to far lower temperatures as discussed in the previous section. An overview of the mirror sector cosmology is shown on the temperature axis on the right-hand side of Fig.~\ref{fig:spectrum}, for $v' = 8 \times 10^7$ GeV, which we justify below. For a sufficiently high reheating temperature after inflation, the following events occur in succession:
\begin{enumerate}  
    \item $t'$, $b'$, $\tau'$, $c'$, $\mu'$, $s'$ and $d'$ freeze-out, and $t'$, $b'$, $\tau'$ and $c'$ subsequently decay. The mirror neutrinos $\nu'$ decouple while relativistic. During this early era, axions keep the SM and SM' sectors in equilibrium since axion-gluon and axion-mirror gluon scatterings are in equilibrium.
    \item The mirror QCD phase transition occurs at $T_{\rm QCD'} \simeq 34$ GeV.  $u'$, $d'$ and the residual frozen-out $s'$ confine to mirror hadrons which quickly annihilate to $\gamma'$ or decay to the lightest stable hadron, $u' u' u'$~\cite{Kang:2006yd,Harigaya:2016nlg,DeLuca:2018mzn}. 
    Any residual quarks containing a PQ component also confine and efficiently annihilate, reducing their abundance far below that of dark matter and thus cosmologically harmless when they eventually decay.
    \item Mirror glueballs ($S'$) also form at the QCD$'$ phase transition and carry most of the energy and entropy of the mirror gluons. $S'$ decays and inverse decays to $a$ are much faster than Hubble, keeping $S'$ in equilibrium to temperatures $T \ll m_{S'}$. The energy and entropy of $S'$ are thus transferred to the Standard and mirror baths which are still in equilibrium with $a$ due to axion-gluon and axion-mirror Primakoff interactions, respectively.
    \item $e'$ freezes-out. The exponential reduction in the number density of $e'$ causes the $\gamma'$ to decouple from $a$ due to the reduction in the axion-mirror Primakoff rate. This marks the decoupling of the Standard and Mirror baths, around $T \sim m_{e'}/10 \sim (20-30)$ GeV (see the blue curve of Fig.~\ref{fig:axionDecoupling}). 
    \item The axion remains in equilibrium with the Standard Model bath due to interactions with gluons and then pions until  $T \sim 10$ MeV.
    \item $\nu'$ and $a'$ decay.
\end{enumerate}
Analytically, we can expect that the critical density of $e'$ will scale as \cite{Kolb:1990vq}
\begin{align}
    \label{eq:omegaDM}
    \Omega_{\rm DM}h^2 \approx 
    \frac{x_{\rm FO}}{M_{\rm Pl} \, \sigma_{e' \bar{e'} \leftrightarrow \gamma' \gamma'}}\frac{\sqrt{g_*(T_{\rm FO}})}{g_{*S}(T_{\rm FO})}\frac{1}{\rm eV} \simeq 0.12 \left(\frac{v'}{8 \times 10^7 \, \rm GeV}\right)^{2}
\end{align}
where $x_{\rm FO} \equiv m_{e'}/{T_{\rm FO}} \simeq 23$ is the ratio of the mirror electron mass to its freeze-out temperature, and $\sigma_{e' \bar{e'} \leftrightarrow \gamma' \gamma'} \simeq \pi \alpha/m_e'^2$ is the mirror electron annihilation cross-section into mirror photons. Because $\Omega_{\rm DM} \propto 1/\sigma_{e' \bar{e'} \leftrightarrow \gamma' \gamma'} \propto m_{e'}^2$, we expect $\Omega_{\rm DM} \propto v'^2$ as shown in Eq.~\eqref{eq:omegaDM}, with $v' \simeq  10^8$ GeV giving $e'$ the observed dark matter abundance. 

In this work, we numerically solve a system of Boltzmann equations describing the cosmology of $e'$ and $u'$ freeze-out in order to precisely determine the $v'$ that gives the correct $e'$ dark matter abundance. We find that Eq.~\eqref{eq:omegaDM} is an excellent approximation and that the relic abundance of $e'$ matches that of the observed dark matter abundance for $v' = 8 \times 10^7$~GeV. This justifies using the value of $v'$ in Fig.~
\ref{fig:spectrum}. In addition, we find that the $u'$ abundance is subdominant compared to $e'$ due to its bound-state formation and thus enhanced annihilations at $T'_{\rm QCD}$. We use the cross-sections as given in Appendix A of \cite{Dunsky:2019upk} to compute this. 

Note that in solving the Boltzmann equations, we require that the $\nu'$ do not dominate the energy density of the Universe before decaying. This condition requires the neutrino Yukawa couplings, $|y_{ij}|$, to be sufficiently large that the $\nu'$ lifetime \eqref{eq:nuPlifetime3} is not exceedingly long. Specifically, to avoid $\nu'$ matter domination, the temperature at which the $\nu'$ decay ($T_{\Gamma, \nu'} \sim \sqrt{M_{\rm Pl} \Gamma_{\nu'}}$) must be greater than the temperature at which they would begin dominating the energy density of the universe ($T_{{\rm MD}, \nu'} \sim m_{\nu'} Y_{\rm therm}$),  
\begin{align}
    \label{eq:dilutionFactor}
    D^{-1} \equiv \frac{T_{\Gamma, \nu'}}{T_{{\rm MD}, \nu'}} = \frac{\sqrt{\Gamma_{\nu'} M_{\rm Pl} \sqrt{\frac{10}{8\pi^3 g_*}}}}{m_{\nu'} Y_{\rm therm}} > 1 \qquad \text{(No $e'$  dilution from $\nu'$ decay)}
\end{align}
for each of the three $\nu'_i$. If $D$ is larger than unity, the $e'$ abundance is diluted by $\nu'$ decay by a factor $D$; for $e'$ to account for all dark matter requires $v' \approx \sqrt{D} \, 10^8$ GeV.

\begin{figure}[tb]
    \centering
    \includegraphics[width=0.9\textwidth]{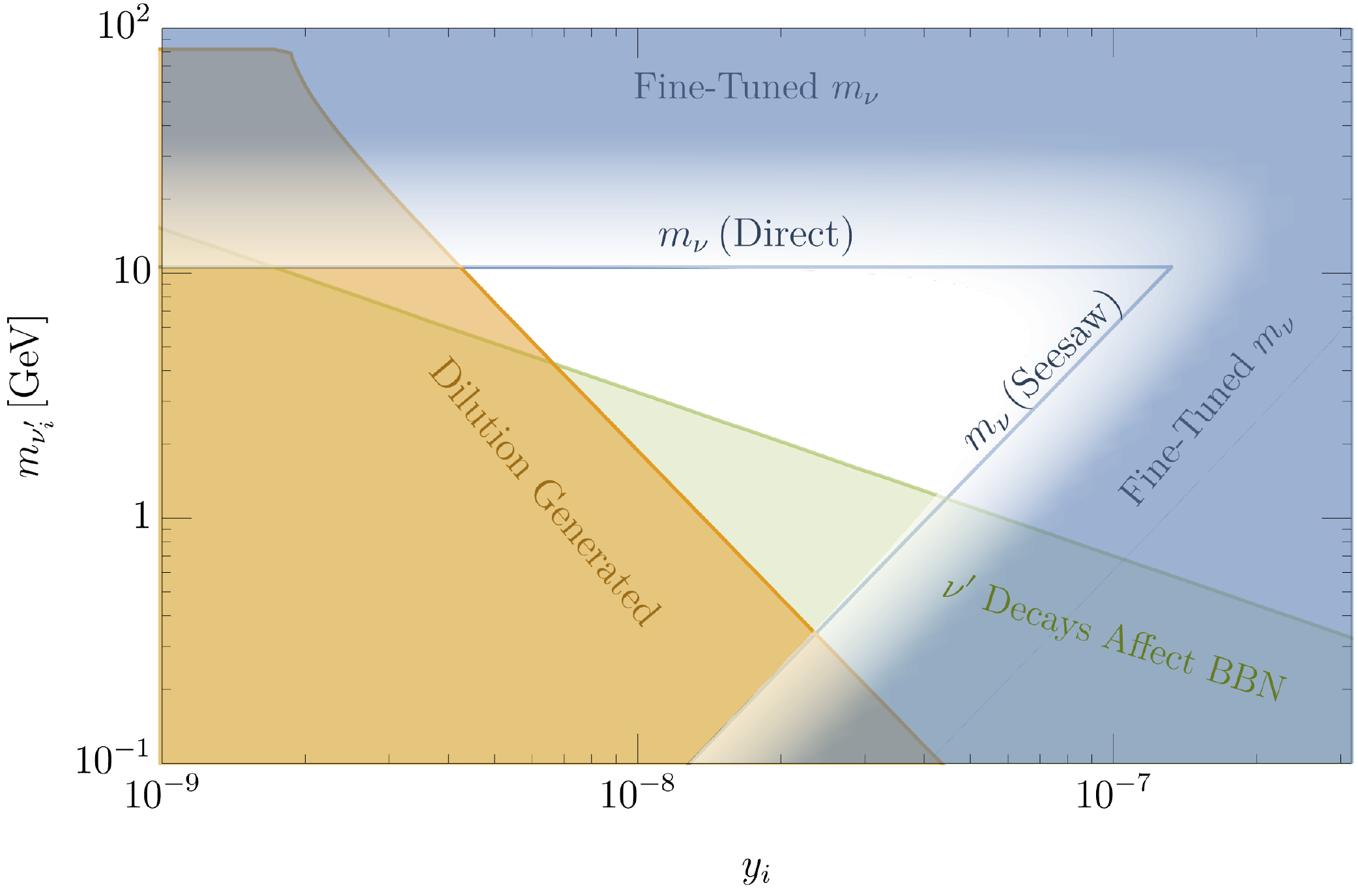} 
    \caption{
    $\nu'$ masses and Yukawa couplings that yield a successful cosmology, with the observed dark matter resulting from $e'$ freeze-out with $v' = 8 \times 10^7$ GeV. 
    The orange region is excluded from $\nu'$ generating entropy upon its decay and diluting the $e'$ abundance below that of dark matter while the green region is excluded from $\nu'$ decaying below $4$ MeV and disrupting nuclear abundances as predicted by standard BBN. The blue region is excluded by naturalness of the $\nu$ masses.}
    \label{fig:zoomedNuPlot}
\end{figure}

In Fig.~\ref{fig:zoomedNuPlot}, the region where $\nu'_i$ decay leads to dilution, $D>1, $ is shaded in orange in the $(y_i, m_{\nu'_i})$ plane. Here an initial thermal abundance of $\nu'$ is assumed, as results from virtual $W'$ exchange if the reheat temperature of the universe after inflation is above $10^6$ GeV, for $v'$ of $10^8$ GeV. Thus, $e'$ freeze-out yields the observed dark matter abundance above and to the right of this region for $v' = 10^8$ GeV. In the orange region, $\nu'_i$ come to dominate the energy density of the universe before they decay; the $e'$ abundance is diluted, requiring larger values of $v'$ to explain the observed dark matter, as studied in section \ref{sec:increasingv'}.

If $\nu'_i$ have significant abundances, they cannot decay during or after Big Bang Nucleosynthesis (BBN) as this leads to unacceptable nuclear abundances \cite{Kawasaki:2004qu}. The green region of Fig.~\ref{fig:zoomedNuPlot} shows the parameter space where $\nu'_i$ decay at temperatures below $T_{\rm BBN} = 4$ MeV.
As the reheat temperature of the universe is lowered below $10^6$ GeV, so that the $\nu'_i$ abundances drop below the thermal one; the orange region shrinks rapidly, but the green region is very robust because the strong BBN bound.  Hence, for a wide range of reheat temperatures, a consistent cosmology for $e'$ dark matter with $v' = 8 \times 10^7$ GeV results only in the region above and to the right of the green region.

On the solid blue contours of Fig.~\ref{fig:zoomedNuPlot}, the direct and seesaw contributions to light neutrino masses, as given in (\ref{eq:Lnu}), are 0.05 eV, so that natural neutrino masses arise in the region between these contours. In the blue shaded regions, unnatural fine tunings must be made in the light neutrino mass matrix; for example a cancellation of a direct mass from one $\nu'$ with the seesaw contribution generated by another $\nu'$.  The beginning of the blue shading marks  the parameter space where the cancellations require a fine tuning by a factor of three.

In summary, for this cosmology to be successful and yield $e'$ dark matter with $v' = 8 \times 10^7$ GeV, all three $\nu'_i$ must lie in the unshaded region of Fig.~\ref{fig:zoomedNuPlot}: in the orange region $e'$ are diluted by a thermal abundance of $\nu'_i$, in the green region late decaying $\nu'_i$ adversely affect BBN for a wide range of $\nu'_i$ abundances, and in the blue region fine-tuning occurs in the neutrino mass matrix.  This implies $m_{\nu'_i} \simeq (0.8-4.0)$ GeV and $y_i \simeq (0.8-4.0) \times 10^{-8}$.  The mirror neutrinos decay well after the electroweak phase transition, so they do not give a baryon asymmetry via leptogenesis.
The light neutrino mass matrix has relevant contributions from three seesaw terms and three direct terms.
The direct masses are all larger than about $0.05 \, \rm{eV} /10$ and the seesaw masses are all larger than about $0.05 \, \rm{eV} /100$. If the reheating temperature of the universe is below $10^6$ GeV, the constraint given by the orange region may be avoided.

While the BBN constraint is quite robust, it is possible to evade if the abundance of $\nu'_i$ is sufficiently low.  This requires the reheat temperature after inflation to be less than about 100 GeV to suppress freeze-in production via virtual $W'$. This is still consistent with $e'$ freeze-out and with gravity wave signals from the QCD$'$ phase transition.  It also requires $y_i \lsim 10^{-15}$ to suppress freeze-in production from the Higgs portal. Thus one or more $\nu'_i$ could lie to the far left of an extension of Fig.~\ref{fig:zoomedNuPlot}, with extremely low $y_i$ and lifetimes longer than $10^9$s.

Of particular interest is the possibility that the neutrinos are Dirac, as discussed in 
section \ref{sec:diracnu}. In this case the heavy $\nu'$ states are absent and hence cannot dilute $e'$ or distort element abundances during BBN; 
Fig.~\ref{fig:zoomedNuPlot} is not applicable.
With $v' = 8\times 10^7$ GeV, mirror electrons account for the observed dark matter for any values of the neutrino masses and mixings consistent with current data. 

Alternatively, neutrino masses may be constrained by separate lepton parities in the SM and mirror sectors, also discussed in section \ref{sec:diracnu}, so that $y_{ij}=0$. This case gives the $y_i=0$ limit of Fig.~\ref{fig:zoomedNuPlot}.  Since there are only direct contributions to the light neutrino masses, the two heavier mirror neutrinos must have masses of order 10 GeV. They have radiative decays, via a virtual loop of $W'$ and $e'$, to a mirror photon and the lightest mirror neutrino, which is stable, with lifetimes of order $10^9$s.  These decays do not affect the BBN abundances or CMB
data, but we must require that $e'$ dark matter is not diluted and the contribution of the lightest $\nu'$ to dark matter is sub-dominant.  This will certainly be the case if the freeze-in abundance of $\nu'$ via virtual $W'$ gives a yield $Y_{\nu'} < 10^{-10}$, which results if the reheat temperature after inflation is less than about 1 TeV. 

\section{Signals}
\label{sec:Signals}
The mirror world, with $v'$ determined to be near $10^8$ GeV by the $e'$ freeze-out abundance, has several observational signals that depend on few free parameters. These signals do not depend on whether $v'$ is set by the Higgs Parity mechanism or by soft $Z_2$ breaking Higgs interactions. Even though this scale is far above that of collider energies -- so that mirror particles are not directly accessible -- a combination of observations could provide strong evidence for the theory. We first discuss signals from dark radiation in Sec. \ref{sec:neff}, and the rare kaon decay $K \rightarrow \pi a$, in Sec. \ref{sec:Kdecays}, which depend on $f_a$. Dark radiation is also sensitive to $E/N$, the ratio of PQ anomalies. Next, we discuss signals from self-interactions of $e'$ in galactic halos in Sec. \ref{sec:selfint} and gravity waves in Sec. \ref{sec:GW}. With $v'$ fixed, these signals do not depend on any unknown parameters of the theory, though there are uncertainties in the computed signals.
\subsection{Dark Radiation, $N_{\rm eff}$}
\label{sec:neff}
\begin{figure}[tb]
    \centering        
    \includegraphics[width=.45\textwidth]{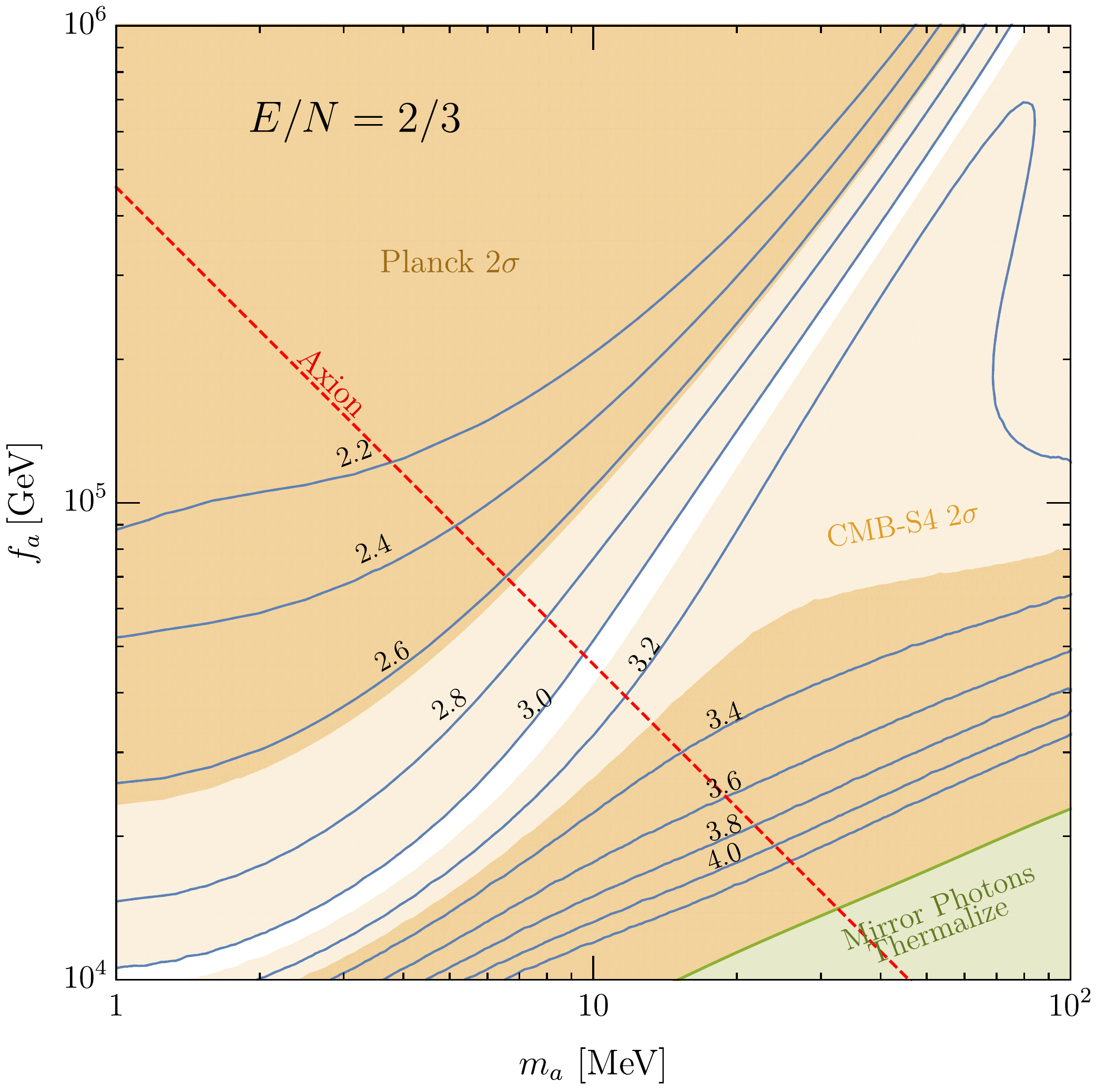} 
        \hfill
        \includegraphics[width=.475\textwidth]{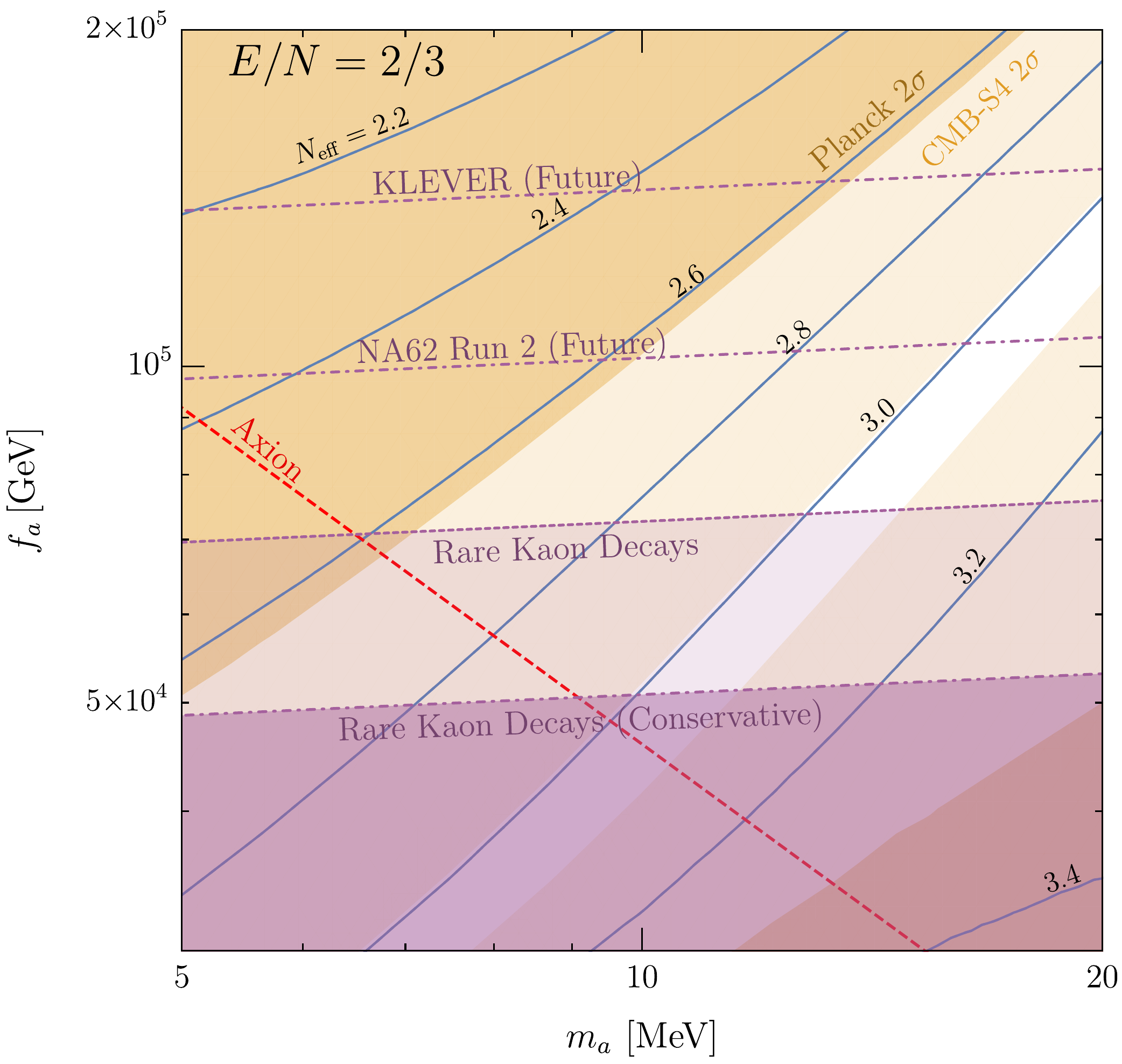}
    \caption{Left: Contours of $N_{\rm eff}$ in the $m_a - f_a$ plane for $E/N = 2/3$ arising from axion decays and the relic mirror photon abundance. The dark orange region is excluded at 95$\%$ confidence by Planck, while the light orange shows the future reach of CMB-S4 experiment at 95$\%$ confidence. The green region shows the parameter space where the mirror photons thermalize and the freeze-in picture we use to compute $N_{\rm eff}$ breaks down. The red dashed line shows the constraint (\ref{eq:ma}) on $m_a f_a$ for $v' = 8 \times 10^7$ GeV. Right: Same as left but zoomed in on the allowed region and showing constraints from rare kaon decays in purple.}
    \label{fig:NeffPlot}
\end{figure}

There are two effects that lead to a change in the dark radiation,  $\Delta N_{\rm eff}$: 1) the decay of the axion around the neutrino decoupling era heats up the photon bath relative to the neutrino bath which generates a \textit{negative} $\Delta N_{\rm eff}$, and 2) the relic mirror photons act as dark radiation which generate a \textit{positive} $\Delta N_{\rm eff}$. Quantitatively, these effect are described by the effective number of neutrinos, 
\begin{align}
    N_{\rm eff} = \frac{8}{7}\left(\frac{11}{4} \right)^{4/3} \frac{\rho_\nu + \rho_{\gamma'}}{\rho_\gamma},
\end{align}
where the expected Standard Model value is $N_{\rm eff}^{(\rm SM)} = 3.044$ \cite{Bennett:2020zkv,Froustey:2020mcq,particle2020review}. In previous work \cite{Dunsky:2022uoq}, we numerically computed the value of $N_{\rm eff}$ for general heavy QCD axion theories with and without frozen-in dark photons. Here we extend that work to incorporate the contribution to $N_{\rm eff}$ from mirror photons frozen-out at the two-sector decoupling temperature of around $\sim 20\mathchar`-30$~GeV. Fig.~\ref{fig:NeffPlot} shows the constraints in the $(m_a, f_a)$ plane for $E/N = 2/3$, where $E/N$ is the ratio of the electric to color anomaly of the PQ quarks when these quarks possess down-quark like quantum numbers.  
\begin{figure}[tb]
    \centering        
    \includegraphics[width=.45\textwidth]{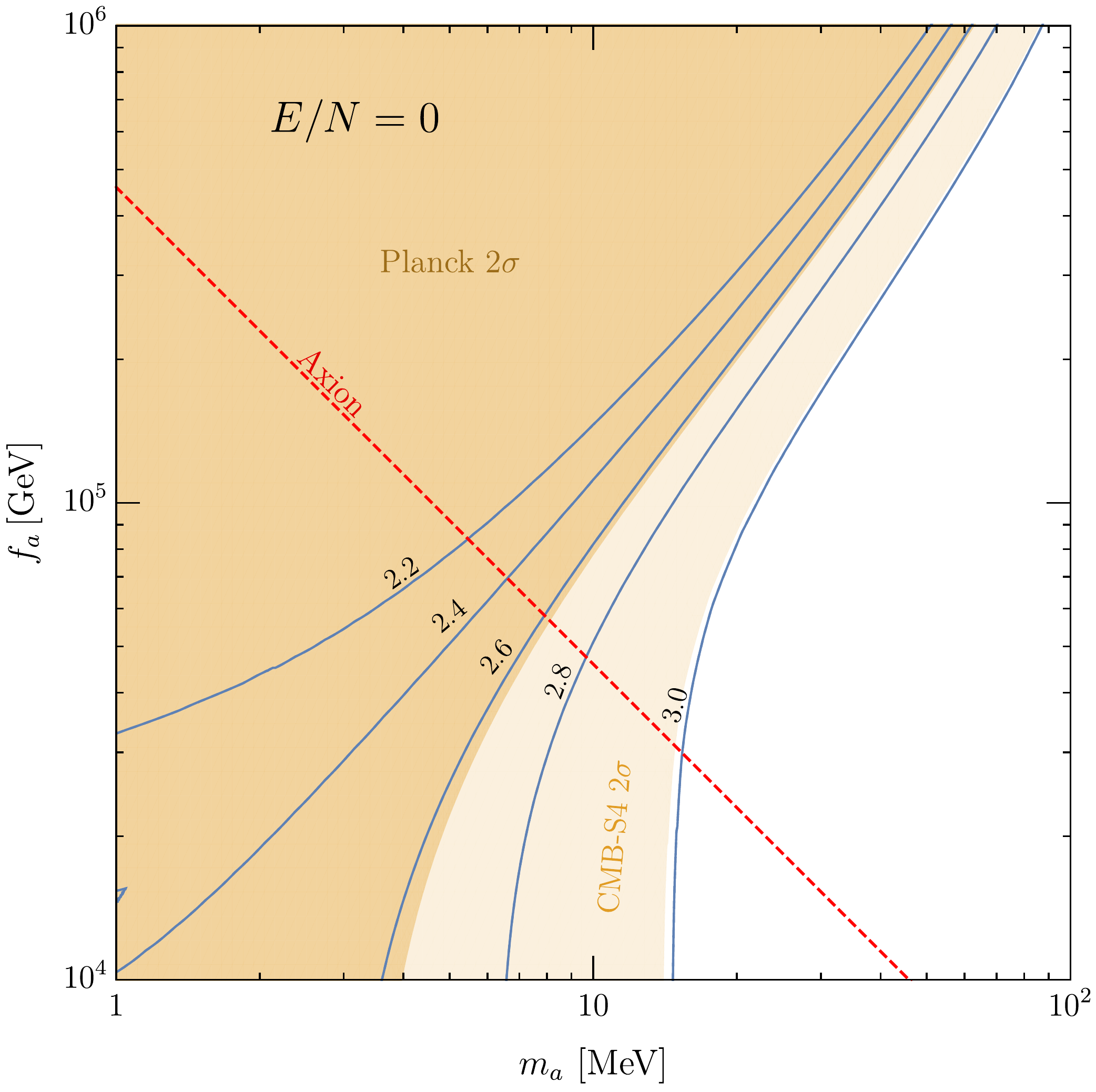} 
        \hfill
        \includegraphics[width=.475\textwidth]{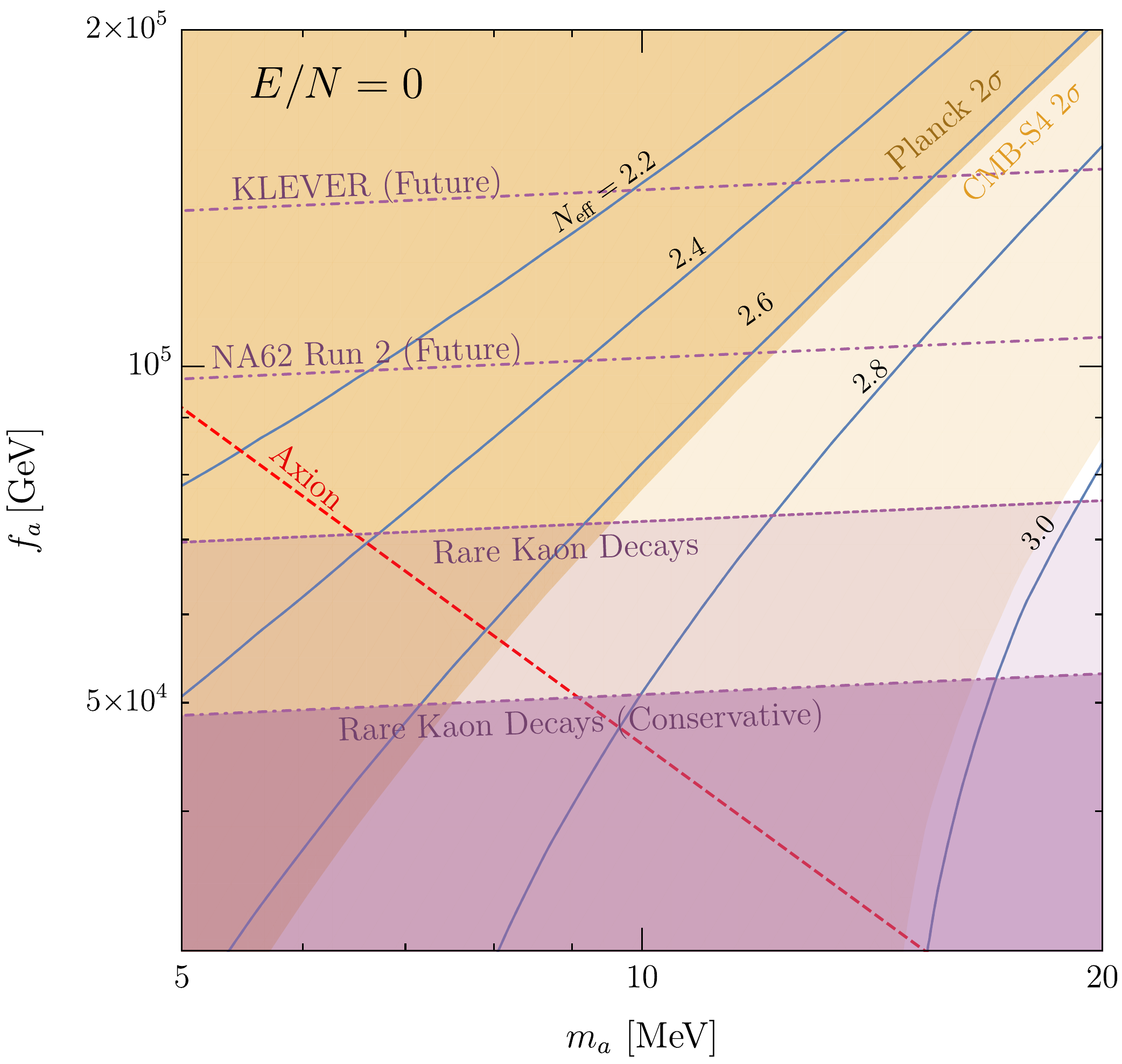}
    \caption{Same as Fig.~\ref{fig:NeffPlot}, but for $E/N = 0$. For $E/N = 0$, the $a \rightarrow 2 \gamma'$ branching ratio is negligible so that $\Delta N_{\rm eff} \simeq 0$ when the axion decays well before neutrino decoupling as shown by the white region in the large $m_a$ and low $f_a$ region of the left panel.}
    \label{fig:NeffPlotEN0}
\end{figure}
The blue contours of Fig.~\ref{fig:NeffPlot} indicate the value of $N_{\rm eff}$ in the $(m_a, f_a)$ plane. The dark orange region shows the current $2\sigma$ limit on $N_{\rm eff}$ from Planck \cite{Planck:2018vyg} while the light orange region shows the future reach of CMB-S4 \cite{CMB-S4:2016ple}. To get the right $e'$ dark matter abundance, the axion mass must lie on the red strip as given by Eq.~\eqref{eq:ma} for $v' = 8 \times 10^7$~GeV. 

We see that most of the axion parameter space is either ruled out by Planck or will be probed by CMB-S4. The intersection of the presently allowed region (light orange) and the locus of points on which the axion can live (dashed red line), powerfully constrains $m_a$ to be between $6.5-13.5$ MeV and $f_a$ between $(3-7) \times 10^4$ GeV. The small white strip in the $(m_a,f_a)$ plane where the $N_{\rm eff}$ signal is so small that CMB-S4 cannot probe may still be able to be probed by rare kaon decays as discussed in the next section.  

The left panel of Fig.~\ref{fig:NeffPlotEN0} shows the same $N_{\rm eff}$ as Fig.~\ref{fig:NeffPlot} but for $E/N = 0$. For this charge-to-color anomaly ratio, the axion coupling to mirror photons vanishes. Consequently, the axion decay to mirror photons does not occur and the only positive contribution to $N_{\rm eff}$ originates in the $\gamma'$ frozen-out at $T \sim 30$ GeV. This contribution is small and for computational simplicity, we neglect it in computing Fig.~\ref{fig:NeffPlotEN0}.  Compared to the case for $E/N = 2/3$, the region of large $m_a$ and small $f_a$ opens up since in this region, the axion decays far before neutrino decoupling and only into Standard Model particles. 

In general, the larger $E/N$ is, the greater the axion branching ratio into mirror photons and the larger the positive contribution to $N_{\rm eff}$ becomes. For $E/N \gtrsim 1$, $g_{\gamma'} > g_\gamma$ and the region of large $m_a$ and low $f_a$ becomes increasingly ruled out by too large $N_{\rm eff}$ from $a \rightarrow 2 \gamma'$ decays. Consequently, we focus on the case $E/N = 2/3$ (PQ quarks with down-like charges) and $E/N = 0$ (KSVZ-like axion). The GUT-motivated value of $E/N = 8/3$ generates such a large positive $\Delta N_{\rm eff}$ that Planck excludes the entire parameter space for $v' = 8 \times 10^7$ GeV.

\subsection{Axion Signal in Kaon Decays}
\label{sec:Kdecays}
Because the axion mixes with the neutral pseudo-scalar mesons $\pi_0$, $\eta$, and $\eta'$, Standard-Model hadronic decays involving such mesons necessarily include axion decays as well. For sufficiently low $f_a$, the decay $K \rightarrow \pi a$, which arises because of  $a-\pi_0$ mixing, can strongly constrain $f_a$ from bounds on rare kaon decays. For example, in the Standard Model, the decay $K^+$ to $\pi^+$ and missing energy is strongly suppressed as it is a flavor changing neutral current process.  The experimental limit on the branching ratio Br($K^+ \rightarrow \pi^+ \nu \bar{\nu})$ is approximately $1 \times 10^{-10}$, which can be translated to a similar bound on the branching ratio Br($K^+ \rightarrow \pi^+ a)$ due to its experimentally similar signature of $\pi^+$ and missing energy \cite{NA62:2021zjw,Bauer:2021wjo,Bauer:2021mvw,Goudzovski:2022vbt}. For $m_a \lesssim 100$ MeV, this translates to a bound on $f_a \gtrsim (5-7) \times 10^4$ GeV.

The right panel of  Fig.~\ref{fig:NeffPlot} shows a zoomed-in region of the left panel but now overlaid with the kaon bounds. The lighter purple region  shows the parameter in the $(m_a,f_a)$ plane where the axion generates a larger than observed branching ratio for the decay of $K^+$ to $\pi^+$ and missing energy in the NA62 experiment, according to Fig.~7 of \cite{Goudzovski:2022vbt}. We caution there is some uncertainty in the matrix element associated with the octet operator that enhances the $K^+ \rightarrow \pi^+ a$ decay rate and show a more conservative NA62 bound in the darker purple region that is weaker by a factor of $\sim 30\%$~\cite{Co:2022aav}, in accordance with the uncertainty of ${\cal O}(m_K^4/m_\rho^4)$ in the matrix element computed at NLO in~\cite{Cirigliano:2011ny}. Future constraints from the NA62 Run 2 and the KLEVER experiment will improve the limit on $f_a$ by a factor of $2$ to $2\sqrt{2}$, respectively \cite{Goudzovski:2022vbt}. The dot-dashed purple lines shows the conservative region enhanced by these factors.  Fig.~\ref{fig:NeffPlot} demonstrates that even with some uncertainty in  Br$(K \rightarrow a \pi)$, the axion bounds from rare kaon decays is complementary to the $N_{\rm eff}$ bounds, probing regions of lower $f_a$ that are not constrained by dark radiation. In particular, the conservative estimate of the bound from $K$ decays implies that $N_{\rm eff}$ is significantly below the SM value; CMB-S4 or the NA62 Run 2 and KLEVER experiments will either discover a signal or exclude this minimal mirror axion theory.

\subsection{Signals from Self-interactions of $e'$ Dark Matter}
\label{sec:selfint}

Mirror electrons and positrons form the dominant component of dark matter, and have self interactions via mirror QED with $\alpha'(\mu = m_e') \simeq 0.0076$.  Galactic halos have a triaxial structure and self-interactions reduce the anisotropy of the DM velocity distribution, reducing the ellipticity of the halos. The measured ellipticity of NGC720 has led to constraints on self-interactions \cite{Feng:2009mn}. Remarkably, for $m_{e'} = 225$ GeV, our predicted value for $\alpha'$ is right on the upper limit    \cite{Agrawal:2016quu}.
As stressed in \cite{Agrawal:2016quu}, these constraints are not strict bounds.  One caveat is that the constraint results from calculations of the time scale for the halo velocity distribution to become isotropic, whereas the measured ellipticity is probing the mass density distribution, which could evolve on a longer time scale. $N$-body simulations could lead to more precise bounds.  Also, the constraint relies on observations of a single galaxy, NGC720, which might have special features, for example low ellipticity; a more robust result would require analysis of further galaxies. Finally, isotropization effects from large self-interactions could be masked by galaxy mergers, which increase ellipticities.

As mirror QED is unbroken, the effects on large scale structure from mirror electron self-interactions results from frequent small angle scattering, rather than rare large angle scattering. Using the results of~\cite{Agrawal:2016quu}, the momentum-transfer cross-section for our $e'$ dark matter is
\begin{align}
    \frac{\sigma_T}{m_{e'}} \simeq 2 \, \frac{\rm{cm}^2}{\rm{g}}\left(\frac{300 \; \rm{km \; s}^{-1}}{v} \right)^4,
\end{align}
and is highly dependent on the speed $v$. Typical speeds in (dwarf galaxies, galaxies, galaxy clusters) are $(30, 300, 1000) \rm{km \; s}^{-1}$. Numerical N-body simulations, including only dark matter, show that on galactic scales a velocity-independent $\sigma_T/m$ of $1 \rm{g} / \rm{cm}^2 $ leads to significant departures from CDM simulations \cite{Fischer:2022rko}.  These departures are seen for several observables, including the matter power spectrum, halo and subhalo mass functions, the number of satellites, halo density and circular velocity profiles, and halo shape distributions.  In particular, for the number of satellites, the density profiles and the shape distributions, the departures from CDM are larger when frequent small-angle scattering dominates over rare large-angle scattering.  This suggests that future numerical simulations, with $\sigma_T \propto 1/v^4$, would identify signals for mirror electron dark matter that could be searched for in the data.

\subsection{Gravitational Waves from $SU(3)'$ Phase Transition}
\label{sec:GW}
For $v' \gtrsim 10^6$ GeV, the mirror QCD scale is below the mass of $u'$, the lightest mirror quark. Consequently, for $v' = 8 \times 10^7$ GeV, the mirror QCD phase transition is first order~\cite{Yaffe:1982qf,Svetitsky:1982gs}. The first-order phase transition creates vacuum bubbles that expand and relativistically collide, generating a gravitational wave spectrum \cite{Witten:1984rs}. The gravitational waves arise from bubble collisions, plasma turbulence, and plasma sound waves \cite{caprini2018cosmological}, so that the spectrum maybe written as
\begin{align}
\label{eq:OmegaGWtot}
\Omega_{\rm GW}h^2 = \left(\Omega_{\rm GW}^{(\rm col)} + \Omega_{\rm GW}^{(\rm turb)} + \Omega_{\rm GW}^{(\rm sound)}\right)h^2 \, .
\end{align}
Two key input quantities that determine the amplitude of the spectrum are the ratio of vacuum energy density to the background radiation density at bubble nucleation,
\begin{align}
    \alpha \equiv \frac{\rho_{\rm vac}}{\rho_{\rm rad}} 
\end{align}
and the ratio of vacuum energy density to the component of the plasma density that strongly couples to the expanding bubble,
\begin{align}
    \alpha' \equiv \frac{\rho_{\rm vac}}{\rho_{\rm coupled}} \, .
\end{align}
These energy density ratios affect the bubble wall velocity \cite{huber2008gravitational},
\begin{align}
    v_b \simeq \frac{\sqrt{\frac{1}{3}} + \sqrt{\alpha'^2 + \frac{2\alpha'}{3}}}{1+\alpha'}  \, ,
\end{align}
the efficiency of the transfer of vacuum energy into bubble kinetic energy \cite{huber2008gravitational}
\begin{align}
    \label{eq:kappa}
    \kappa \equiv \frac{\rho_{\rm kin}}{\rho_{\rm vac}} \simeq \frac{1}{1+0.715 \alpha'}\left(0.715 \alpha' + \frac{4}{27}\sqrt{\frac{3\alpha'}{2}} \right) \,
\end{align}
and the fractional kinetic energy density of the bubbles to the total energy density of the universe \cite{Caprini:2010xv}
\begin{align}
   \frac{\rho_{\rm kin}}{\rho_{\rm rad} + \rho_{\rm vac}} = \kappa \frac{\alpha}{1+ \alpha} \, . 
\end{align}
Assuming that the energy density of the mirror gluon gas at $T_{\rm QCD'}$ ($\rho_g' \simeq 1.07T_{\rm QCD'}^4$ \cite{Borsanyi:2012ve}) is comparable to the vacuum energy, then $\alpha \approx 0.03$. The precise value of $\alpha'$ is not as important since the spectrum is fairly insensitive to $\alpha'$ as long as it is $\mathcal{O}(1)$. For a fiducial value, we take $\alpha' \simeq 0.343$ \cite{Morgante:2022zvc}. Last, we assume that the efficiency factor is the same for collisions, turbulence, and sound waves. Other numerical fits of the efficiency factor, such as in \cite{Espinosa:2010hh}, give very similar values to Eq. \eqref{eq:kappa} in the $\alpha' \sim 1$ limit.

The gravitational-wave spectra due to bubble collisions \cite{huber2008gravitational,Caprini:2010xv}, turbulence \cite{Caprini:2009yp,Caprini:2010xv}, and sound waves \cite{Hindmarsh:2015qta,Hindmarsh:2017gnf,Caprini:2019egz} are
\begin{align}
\Omega_{\rm GW}^{(\rm col)}h^2 \simeq \; & \Omega_{\rm rad} h^2  \left(\frac{\kappa \alpha}{1+\alpha} \right)^2 \left(\frac{H}{\beta}\right)^{2}  S_c(f)
\\
\Omega_{\rm GW}^{(\rm turb)}h^2 \simeq \; & \Omega_{\rm rad} h^2  \left(\frac{\kappa \alpha}{1+\alpha} \right)^{3/2} \left(\frac{H}{\beta}\right)  S_t(f)
\\
\Omega_{\rm GW}^{(\rm sound)}h^2 \simeq \; & \Omega_{\rm rad} h^2  \left(\frac{\kappa \alpha}{1+\alpha} \right)^{3/2} \left(\frac{H}{\beta}\right)^2 S_s(f) 
\end{align}
where $\Omega_{\rm rad} h^2 \simeq 4.16 \times 10^{-5}$ is the fractional density of radiation today and $\beta/H$ quantifies the duration of the phase transition. Note that the scaling of the efficiency in $\Omega_{\rm GW}^{(\rm sound)}$ changes for long duration sources which are not applicable in the large $\beta/H$ limit that we take here \cite{Caprini:2019egz}. The frequency dependence and effect of the bubble wall velocity is encoded in the spectral shape functions $S(f)$, which are given for bubble collisions, turbulence, and sound waves by
\begin{align}
    S_c &= \frac{2}{3\pi^2} \frac{v_b^3}{0.42 + v_b^2}  \frac{(f/f_c)^3}{1/3+ (f/f_c)^4} 
    \\
    S_t &= \frac{2}{\pi^2} v_b^2 \frac{(f/f_t)^3 (1 + f/f_t)^{-11/3}}{(f/f_t)(\beta/H) + v_b/4\pi^2}
    \\
    S_s &= 2.4 \times 10^{-2} \, v_b^2 (f/f_s)^3 \left(\frac{7}{4+3(f/f_s)^2}\right)^{7/2} \, ,
\end{align}
with reference frequencies
\begin{align}
    f_c &= 3.5 \times 10^{-6} \, {\rm Hz} \left(\frac{\beta}{H}\right)\left(\frac{T_{\rm QCD'}}{35 \, \rm GeV}\right) \left( \frac{g_*}{100}\right)^{1/6} \left(\frac{1}{1.8-0.1v_b + v_b^2}\right)
    \\
    f_t &= 8.5 \times 10^{-6} \, {\rm Hz} \left(\frac{\beta}{H}\right)\left(\frac{T_{\rm QCD'}}{35 \, \rm GeV}\right) \left( \frac{g_*}{100}\right)^{1/6} \left(\frac{1}{v_b}\right) 
    \\
    f_s &= 3.0 \times 10^{-6}  \, {\rm Hz} \left(\frac{\beta}{H}\right)\left(\frac{T_{\rm QCD'}}{35 \, \rm GeV}\right) \left( \frac{g_*}{100}\right)^{1/6} \left(\frac{1}{v_b}\right)  \, .
\end{align}
In this work, the temperature of the phase transition, $T_{\rm QCD'}$, is determined by (\ref{eq:TQCD'}) with $v' \simeq 8 \times 10^7$ GeV in order for the relic freeze-out abundance of $e'$ to match the observed dark matter abundance. As given by Eq.~\eqref{eq:mirrorQCDscale}, this sets $\Lambda_{\rm QCD'} \simeq 27$ GeV, or equivalently, $T_{\rm QCD'} \simeq 34$ GeV. The duration of the phase transition, $\beta^{-1}$, is the main unknown parameter that enters into the gravitational wave spectrum. Typically, the duration of the phase transition is expected to be some fraction of Hubble, $H$, with $\beta/H$ between $100-10^5$. Here 100 is the maximal possible value~\cite{Hogan:1983ixn}, while $10^5$ is a value suggested by a holographic computation~\cite{Morgante:2022zvc}. Note the larger value of $\beta/H$ (i.e., a shorter duration phase transition) gives rise to a smaller amplitude in $\Omega_{\rm GW}h^2$ but also a higher peak frequency of $f_c$ or $f_t$.

\begin{figure}[tb]
    \centering
    \includegraphics[width=1\textwidth]{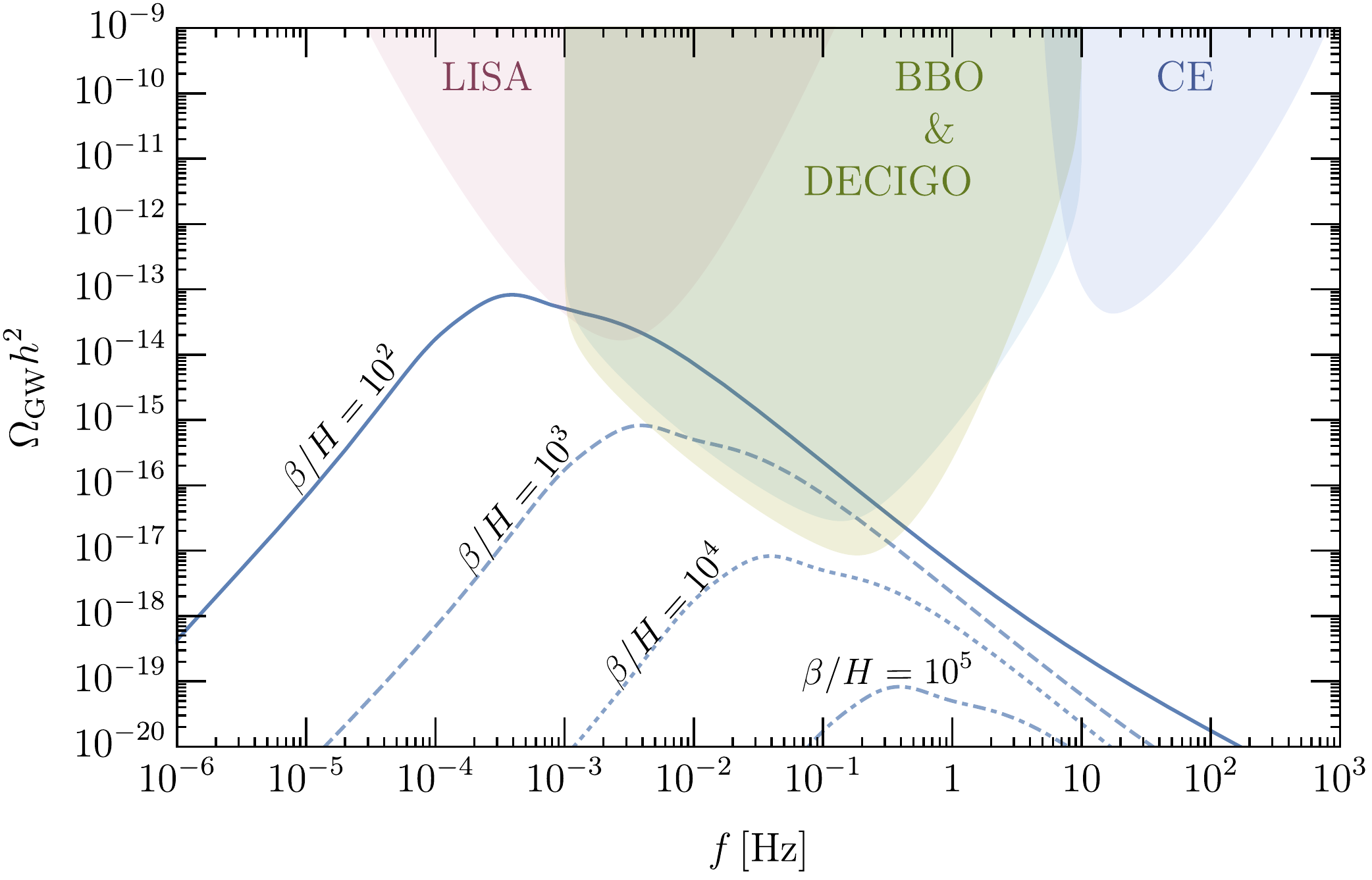} 
    \caption{Gravitational wave signal arising from the first-order mirror QCD phase transition. The solid, dashed, dotted, and dot-dashed curves show $\Omega_{\rm GW} h^2$ for different assumptions of $\beta/H$. For $\beta/H \lesssim  10^4$, the gravitational wave signal may be detected by the BBO and DECIGO experiments.}
    \label{fig:GWPlot}
\end{figure}

$\Omega_{\rm GW}h^2$ is shown in Fig.~\ref{fig:GWPlot} for $\Lambda_{\rm QCD'} \simeq 27$ GeV ($T_{\rm QCD'} \simeq 34$ GeV) and a range of potential values for $\beta/H$. We see that future gravitational wave detectors like LISA, BBO, and DECIGO \cite{amaro2017laser,harry2006laser,Kawamura:2020pcg,Schmitz:2020syl} may be able to observe a gravitational wave signal from the mirror QCD phase transition if $\beta/H$ is not too large.
Once $\beta/H$ is known more precisely from simulations and theory, the amplitude and peak frequency of the gravitational wave signal will be predicted which can provide strong evidence for the simple model of high quality axions in this work.

\section{Increasing the Mirror Electroweak Scale}
\label{sec:increasingv'}
\begin{figure}[tb]
    \centering
    \includegraphics[width=0.9\textwidth]{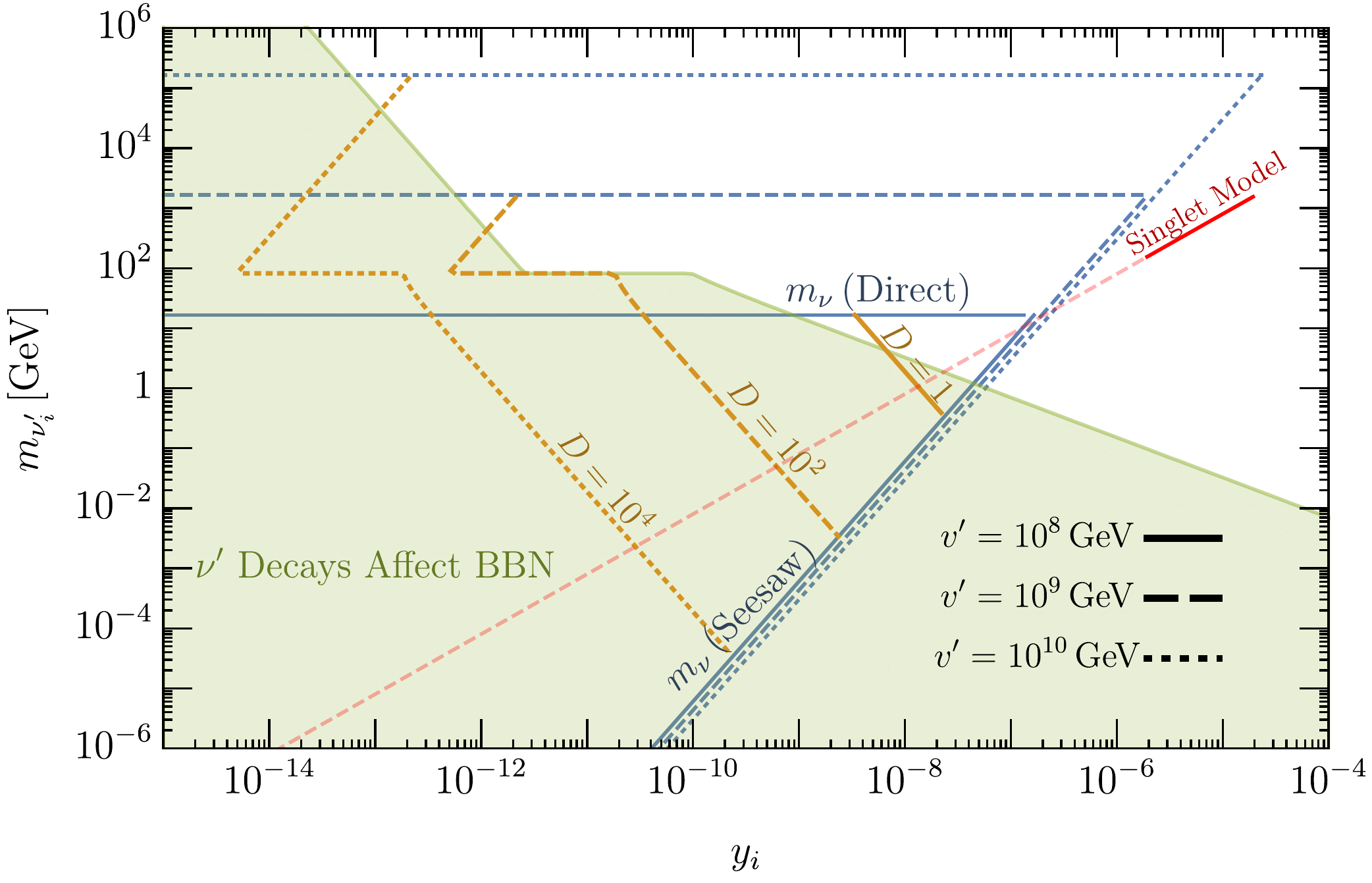} 
    \caption{Generalization of Fig.~\ref{fig:zoomedNuPlot} but for $v' \geq 10^8$ GeV. The solid, dashed, and dotted contours correspond to dilution factors of $1,10^2$ and $10^4$ generated by the decay of $\nu'$. For $v' = 10^8$ GeV, no dilution is required and the allowed parameter space must lie to the right of the solid orange $D = 1$ contour, and for natural $\nu$ masses, within the solid blue triangle. For $v' =10^9  \, (10^{10})$ GeV, $D = 10^2 \,(10^4)$ is required and at least one neutrino mass must lie \textit{on} the dashed (dotted) orange contour and for natural $\nu$ masses, within the dashed (dotted) blue triangle. The blue solid, dashed, and dotted seesaw contributions are artificially displaced slightly to make the allowed natural triangular region for each $v'$ easier to see. The dashed red contour shows the allowed $(m_{\nu'_i}, y_i)$ for the singlet model as discussed in Sec. \ref{sec:singlet}.}
    \label{fig:3genDilution}
\end{figure}
In this paper, we have discussed the cosmology and phenomenological signals of a heavy QCD axion arising from a mirror world. To recap, the logic proceeded as follows: matching the freeze-out abundance of mirror electrons to the observed dark matter abundance fixed $v' \simeq 8 \times 10^7$ GeV. Knowing $v'$, we determined the mirror QCD scale from Eq.~\eqref{eq:mirrorQCDscale}, the axion mass as a function of $f_a$ \eqref{eq:ma}, the mirror neutrino masses  \eqref{eq:numassWeinberg}, and the gravity wave spectrum (Fig.~\ref{fig:GWPlot}). A question naturally arises: if we relax the constraint that the mirror electron is set by pure freeze-out, can $v'$ then be increased above $10^8$ GeV? If so, how does this change the observable signals?

First, with larger $v'$, the parameter space where the neutrino masses are not fined-tuned grows beyond the triangular region of Fig.~\ref{fig:mnuDilutionPlot}. In general, each $\nu'_i$ (with $i = 1,2,3$) leads to two contributions to the light neutrino mass matrix. As shown in (\ref{eq:Lnu}) there is a direct mass term  $(v/v')^2m_{\nu',i}$ for $\nu_i$ and a seesaw term $y_i^2 v^2/m_{\nu',i}$ for the active neutrino $\tilde{\nu}_i$, defined by (\ref{eq:Lyi}). Each term can naturally be at most $\sim 0.05$ eV (that is, without any fine-tuned cancellation). The horizontal `Direct' branch contours in Fig.~\ref{fig:3genDilution} indicate where the  direct mass term is 0.05 eV for $v' = 10^8$ GeV (solid), $10^9$ GeV (dashed), and $10^{10}$ GeV (dotted), which grows quadratically with $v'$. Similarly, the diagonal `Seesaw' branch contours in Fig.~\ref{fig:3genDilution} indicate where the seesaw mass term is 0.05 eV for the same three values of $v'$. Note it is possible that one light neutrino is much less than $0.05$ eV, in which case the allowed parameter space lies \textit{within} the triangle bounded by the Direct and Seesaw contours. However, at least one light neutrino should have a mass comparable to $0.05$ eV in which case at least one $m_{\nu',i}$ lives approximately on the Direct or Seesaw contour. 

Second, when $v'$ increases, $m_e'$ increases so that matching the relic yield of $e'$ to the observable dark matter energy density requires that $e'$ possesses a yield lower than that from freeze-out. One way to realize this is to dilute the $e'$ abundance after it freezes-out, which can be accomplished naturally by long-lived $\nu'$. The orange contours of Fig.~\ref{fig:3genDilution} show different values of the dilution factor (see Eq.~ \eqref{eq:dilutionFactor}) in the $y_{\nu}-m_\nu'$ plane. To the right of the solid orange ($D = 1$) line, $y_{\nu}$ is sufficiently large that $\nu'$ decays early enough to avoid generating any dilution. This is the region for $v' = 10^8$ GeV, where no dilution of $e'$ is necessary, and is shown enlarged in Fig.~\ref{fig:zoomedNuPlot}. To the left of this line, $e'$ dilution from $\nu'$ decay occurs. Since the frozen-out density of $e'$ scales as $\Omega_{\rm e'} \propto m_{e'}^2 \propto v'^2$, and since no dilution of $e'$ is required for $v' = 10^8$ GeV, it follows that the required $e'$ dilution as a function of $v'$ is $D \approx (v'/10^8 \, {\rm GeV})^2$. The dashed and dotted orange contours show the parameter space where $D = 10^2$ and $10^4$, respectively. To achieve the correct dilution for $v' = 10^9$ GeV  and $10^{10}$ GeV requires that one of the $\nu'_i$ lives on the dashed or dotted orange contours, respectively, and none live to the left.

This case of $e'$ dark matter in the mirror world, via freeze-out and dilution from $\nu'$ decays, was considered in \cite{Dunsky:2019upk}. In that paper there was no axion, so $y_D$ of (\ref{eq:LYD}) was absent and $v'$ was determined by the Higgs Parity mechanism, with no soft breaking of the $Z_2$ and with only SM interactions responsible for the radiative generation of the SM quartic. Fig.~10 of \cite{Dunsky:2019upk} then shows that $e'$ dark matter requires $v' \lsim 3 \times 10^{10}$ GeV. Such $v'$ do not result from the central experimental values of $m_t = (172.56 \pm 0.48)$ GeV and $\alpha_s(M_Z) = 0.1179 \pm 0.0009$. For example, for central $\alpha_s$ they require $m_t$ to be at least 2$\sigma$ high; alternatively if $m_t$ is 1$\sigma$ high then $\alpha_s$ must be more than 1$\sigma$ low. More precise measurements of $m_t$ and $\alpha_s$ will determine if this scheme of \cite{Dunsky:2019upk} is allowed.

Third, to get the necessary dilution, $\nu'$ cannot decay after BBN without distorting heavy element abundances. The green region of Fig.~\ref{fig:3genDilution} shows the parameter space where $\nu'$ decays at temperatures below $T_{\rm BBN} = 4$ MeV.
Fig.~\ref{fig:3genDilution} demonstrates that the necessary dilution for $e'$ dark matter, $D = 10^2$ for $v' = 10^{9}$ GeV ($D = 10^4$ for $v' = 10^{10}$ GeV), requires one $\nu'_i$ to live on the short segment of the orange dashed (dotted) line in the unshaded region. The length of this line segment is constrained by avoiding fine-tuning in the neutrino sector and requiring $\nu'$ to decay before BBN. This $\nu'_i$, with a mass of order $10^3$ GeV ($10^5$ GeV), gives a relevant direct mass contribution to the corresponding light neutrino.  The other two $\nu'_i$ can lie anywhere to the right of the orange dashed (dotted) line (although still within the blue dashed or dotted lines for naturalness), and at least one of them must give another relevant contribution to the light neutrino masses.  

Fourth, we emphasize that while allowing $\nu'$ to dilute the $e'$ abundance makes $v' > 10^8$ GeV feasible, it does so at the cost of diminishing some cosmological signals. For example, the gravitational wave spectrum shown in Fig.~\ref{fig:GWPlot} is reduced by a factor of $D^{4/3}$, which can weaken the signal below the detection threshold of LISA, BBO and DECIGO except for the largest $\beta/H$. Similarly, the dilution generated by $\nu'$, which is generated close to BBN, erases most of the $\Delta N_{\rm eff}$ generated by axion decays or residual dark photon radiation shown in Fig.~\ref{fig:NeffPlot}. 

\begin{figure}[tb]
    \centering
        \includegraphics[width=.75\textwidth]{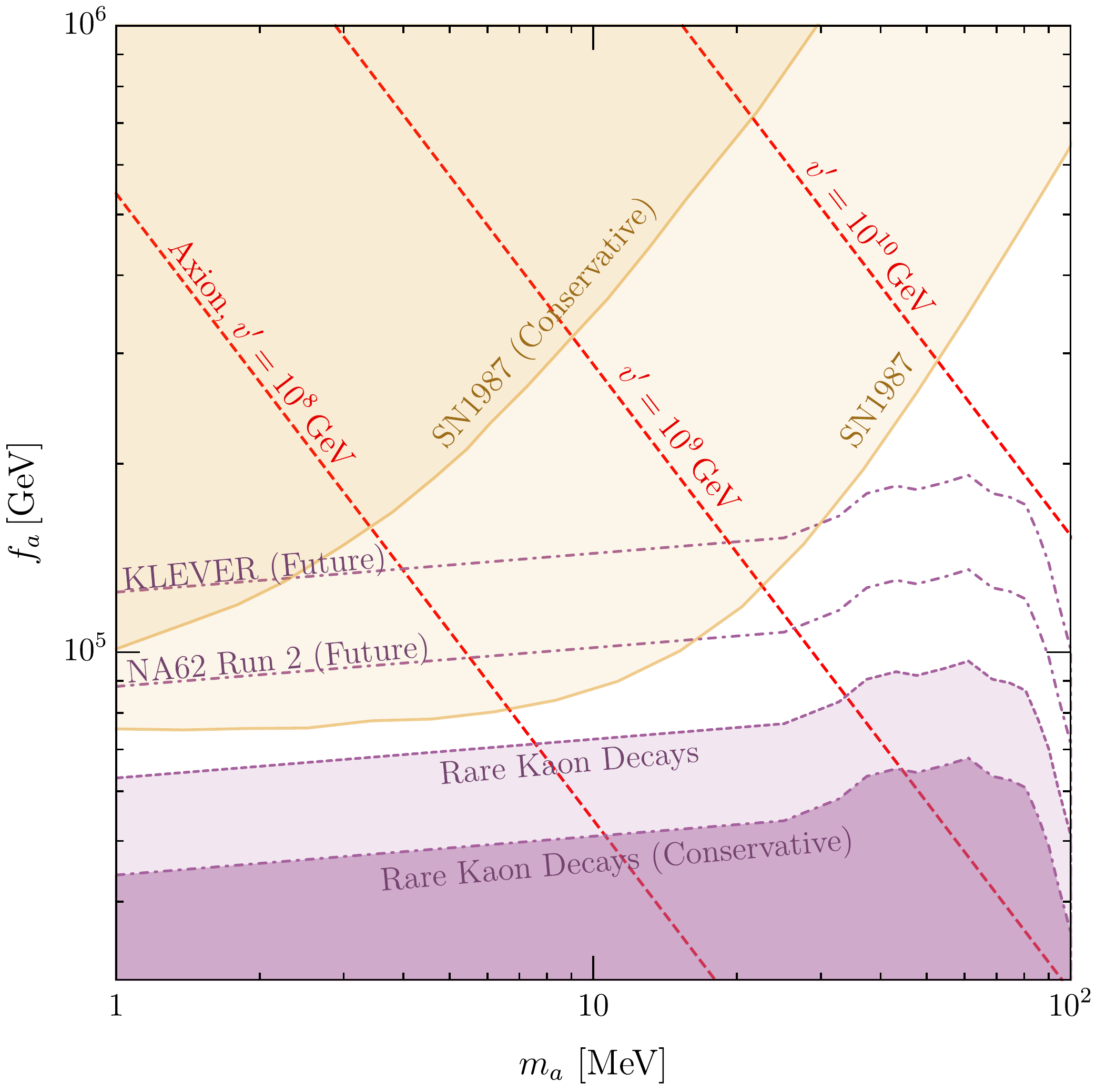}
    \caption{
    For $v' > 10^8$ GeV, dilution of $e'$ dark matter by $\nu'$ decays is required, significantly altering the $N_{\rm eff}$ bounds of Fig.~\ref{fig:NeffPlot}. Nevertheless, supernova (orange) and rare kaon decay (purple) constraints remain. The dashed red contours show the allowed axion mass for $v' = 10^8, 10^9$ and $10^{10}$ GeV. 
    }
    \label{fig:kaonSNPlot}
\end{figure}

Fifth, increasing $v'$ increases $\Lambda_{\rm QCD}'$ so that the axion mass is also heavier according to Eq.~\eqref{eq:ma}. The three dashed red lines of Fig.~\ref{fig:kaonSNPlot} shows the $m_a-f_a$ correlation for $v' = (10^8, 10^9, 10^{10})$ GeV. The  orange regions shows the constraints from SN 1987A summarized in~\cite{Kelly:2020dda} with the darker region~\cite{Chang:2018rso} more conservative for low $f_a$ than the lighter region~\cite{Ertas:2020xcc}. The purple region shows the bound from searches for the rare kaon decay $K^+ \rightarrow \pi^+ a$. 

Sixth, since the constraints from $\Delta N_{\rm eff}$ become less powerful with increasing $v'$, larger values of the PQ anomaly ratio, $E/N$, are allowed. Thus, unlike at $v' = 10^8$ GeV, it is possible for the SM to be unified, for example into an $SU(5)$ theory with $E/N = 8/3$. This introduces two more observational signals that have event rates with very different dependencies on the unified scale $v_G$. The proton decay event rate scales like $1/v_G^4$, while the event rate for the direct detection of $e'$ dark matter from kinetic mixing scales as a positive of $v_G$ (because $\epsilon$ is generated by inserting unified symmetry breaking into a higher dimension operator).  Results for these signals are shown in Figure 5 of \cite{Dunsky:2019upk}.  With $\epsilon$ arising from a dimension 8 operator, proton decay will be discovered at the Hyper-Kamiokande experiment and dark matter at the LZ experiment.

Finally, what if the neutrinos are Dirac, or a lepton parity sets $y_i = 0$, as discussed in section \ref{sec:diracnu}? In the case of Dirac neutrinos, $\nu'$ become $\nu_R$, degenerate with $\nu_L$.  Since there are no heavy $\nu'$ states, it is not possible to dilute the mirror electron abundance by $\nu'$ decay; dark matter requires $v' = 10^8$ GeV. 
For the case with $y_i = 0$ from lepton parity, the two heavier $\nu'$ decay radiatively to the mirror photon and the lightest $\nu'$, which is stable. To give $e'$ dilution and successful nucleosynthesis, these decays must occur before 4 MeV, requiring much larger $m'_\nu$ and therefore larger $v'$ of at least $10^{10}$ GeV. Any further increase in $v'$ makes $\nu'$ sufficiently heavy that the mirror beta decay $\nu' \rightarrow e' u' \bar{d}'$ opens up. The resulting rapid decays of $\nu'$ imply that, at these large values of $v'$, the freeze-out abundance of $e'$ cannot be sufficiently  
diluted.  Hence, this scheme works, if at all, only for a narrow range of $v'$ near $3 \times 10^{10}$ GeV.
\section{Summary}
In this paper, we studied a mirror-world scenario with a PQ symmetry, where the strong CP problem is solved by an axion. The mirror QCD scale is much larger than the SM QCD scale because of the large mirror electroweak scale, $v' \gg v$. The axion mass is dominantly given by mirror QCD dynamics and is larger than the standard case. The axion decay constant may be much below $10^9$ GeV. Because of the larger axion mass and the smaller decay constant, the PQ symmetry may be easily understood as an accidental symmetry.

The mirror electron is absolutely stable because of mirror electromagnetic charge conservation and is a natural dark matter candidate.  Self-interactions from mirror QED affect large scale structure, such as galactic density profiles and shape distributions, and the number of galactic satellites.
The observed dark matter abundance is explained by freeze out of the annihilation of mirror electrons if the mirror electroweak scale $v'$ is around $10^8$ GeV. The Higgs Parity mechanism can set such a value of $v'$ if the coupling of the Higgs with the KSVZ fermions is $O(1)$.
The mirror QCD scale is uniquely predicted, and the axion mass is predicted as a function of the decay constant, as shown in Eq.~\eqref{eq:ma}. The mirror QCD phase transition is first order and produces primordial gravitational waves. Since the unique parameter of the phase transition, namely, the mirror QCD scale, is fixed the spectrum of the gravitational waves can in principle be uniquely predicted. Practically, the mirror QCD phase transition is induced by strong dynamics so we cannot precisely compute the spectrum. We show the spectrum as a function of parameters that characterize the phase transition in Fig.~\ref{fig:GWPlot}.

The heavy axion plays an important role in the cosmology of the mirror world. Without the axion, the mirror QCD entropy would be transferred into mirror photons, producing too much dark radiation. The axion instead maintains the thermal equilibrium of the SM and mirror sectors and the mirror QCD entropy is shared with the SM bath. Still, the radiation energy density of the universe is modified by the presence of mirror photons and by the decay of the axion into SM and mirror photons after neutrinos decouple. The constraint on the parameter space is shown in Figs.~\ref{fig:NeffPlot} and \ref{fig:NeffPlotEN0}. Future observations of the cosmic microwave background and rare kaon decays will probe all of the remaining parameter space.

The mirror neutrinos also affect the cosmological evolution. If they obtain Majorana masses from dimension-five operators, they are massive and may decay after they dominate the universe, diluting mirror electron dark matter and gravitational waves.
This is avoided if the observed neutrino masses are explained by comparable contributions from the dimension-five operators and the see-saw mechanism from the mirror neutrinos, as shown in Fig.~\ref{fig:3genDilution}. Such comparable contributions are a prediction of the minimal UV completion of the dimension-five operators.
If significant entropy production does occur, $v' > 10^8$ GeV becomes also viable. 
If lepton symmetry is preserved, the mirror neutrinos obtain Dirac masses with the SM neutrinos. They are nearly massless and behave as dark radiation.

\begin{acknowledgments}

DD is supported by the James Arthur Postdoctoral Fellowship. The work of LH was supported by the Director, Office of Science, Office of High Energy Physics of the U.S. Department of Energy under the Contract No. DE-AC02-05CH11231 and by the NSF grants PHY-1915314 and PHY-2210390. 
\end{acknowledgments}

\appendix 
\section{Accidental PQ symmetry}
\label{sec:accidental}

In this appendix, we discuss a UV completion where the PQ symmetry accidentally arises as a result of other exact symmetry. We impose $Z_2\times Z_3$ symmetry and introduce a complex scalar field $P$ and three pairs of quarks $D_i$, $\bar{D}_i$ and their mirror partners $D'_i$, $\bar{D}'_i$. Under the $Z_2 \times Z_3$ symmetry, they transform as
\begin{align}
Z_2 &: P\rightarrow -P,~~D_i \rightarrow - D_i',~~\bar{D}_i \rightarrow \bar{D}_i' \nonumber \\
Z_3 & : P\rightarrow  e^{i2\pi/3} \times P,~~ D_i\rightarrow  e^{-i2\pi/3} \times D_i,~~D'_i\rightarrow  e^{-i2\pi/3} \times D'_i
\end{align}
$Z_3\times Z_2$ does not have QCD or gravitational anomaly.
Instead of adding three pairs of fermions in the fundamental representation of $SU(3)$, we may consider higher representations of $SU(3)\times SU(2)$, such as $({\bf 8}, {\bf 1})$  and $({\bf 3}, {\bf 3}) + ({\bf \bar{3}}, {\bf 3}) $. For the latter case, $E/N$ becomes large so we need dilution after the mirror photon decouples or more fermions to reduce the electromagnetic anomaly coefficient.

The renormalizable couplings among them allowed by the symmetry are
\begin{align}
\xi_i P \left(D_i \bar{D}_i +D'_i \bar{D}'_i\right)  - \lambda \left(|P|^2 - \frac{9}{2}f_a^2\right)^2,
\end{align}
which preserve an accidental $U(1)_{\rm PQ}$ symmetry. Here the factor of $9$ comes from the square of the domain wall number.
The leading explicit PQ-breaking term is $P^6/M^2$, where $M$ is the cutoff scale. As long as
\begin{align}
m_a \gtrsim 10^5 \frac{f_a^2}{M} \sim 0.1~{\rm MeV} \left(\frac{f_a}{10^5~{\rm GeV}}\right)^2 \frac{M_{\rm pl}}{M},
\end{align}
the strong CP problem is solved.

It is intriguing to identify the $Z_2$ symmetry above with the mirror symmetry. However, for that case, the leading explicit PQ breaking term is $P^3(|H|^2-|H'|^2)/M$, and the strong CP problem cannot be solved. This can be avoided by embedding the theory into solutions to the electroweak hierarchy problem where $|H|^2$ and $|H'|^2$ are controlled by additional structure, such as supersymmetry or compositeness of the Higgs. 
Another possibility is to generalize the model into the ones with $Z_{2n+1}\times Z_2$ symmetry with $n>1$ and forbid  explicit PQ breaking up to $P^{(2n+1)}(|H|^2- |H'|^2)$. This can be achieved by introducing $2n+1$ pairs of fermions.

\bibliographystyle{JHEP}
\bibliography{biblio}
\end{document}